\documentclass[12pt]{article}
\usepackage{epsfig}
\usepackage{amsmath}
\usepackage{amsfonts}
\usepackage{amssymb}
\usepackage{subfigure}
\usepackage{cite,graphicx}

\newcommand{\abs}[1]{| #1 |}
\newcommand{\ir}{\text{\tiny IR}}
\newcommand{\uv}{\text{\tiny UV}}

\newcommand{\soft}{\text{\tiny soft}}

\newcommand{\amsb}{\text{\tiny amsb}}

\newcommand{\sig}{k |y|}
\newcommand{\be}{\begin{equation}}
\newcommand{\ee}{\end{equation}}

\DeclareMathOperator{\ima}{Im}

\newcommand{\susyb}{\text{\tiny SUSY}}

\newcommand{\kk}{\text{\tiny KK}}
\newcommand{\ew}{\text{\tiny EW}}
\newcommand{\eff}{\text{\tiny eff}}
\newcommand{\scalar}{\text{\tiny scalar}}
\DeclareMathOperator{\Tr}{Tr}

\begin{document}
\pagestyle{empty}

\begin{center}

{\LARGE \bf  A natural little hierarchy for RS \\from accidental SUSY}

\vspace{1.0cm}

{\sc \small Tony Gherghetta}\footnote{E-mail:  tgher@unimelb.edu.au}$^{,a,b}$,
{\sc \small Benedict von Harling}\footnote{E-mail:  bvo@unimelb.edu.au}$^{,a}$
{\footnotesize\sc and }
{\sc \small Nicholas Setzer}\footnote{E-mail:  nsetzer@unimelb.edu.au}$^{,a}$ \\
\vspace{.5cm}
{\it \footnotesize {$^a$School of Physics, University of Melbourne, Victoria 3010,
Australia}}\\
{\it\footnotesize {$^b$Stanford Institute of Theoretical Physics, Stanford University, Stanford, CA 94305,
USA}}
\end{center}

\vspace{1cm}
\begin{abstract}
We use supersymmetry to address the little hierarchy problem in Randall-Sundrum models
by naturally generating a hierarchy between the IR scale and the electroweak scale. 
Supersymmetry is broken on the UV brane which triggers the stabilization of the warped extra 
dimension at an IR scale of order 10 TeV. 
The Higgs and top quark live near the IR brane whereas light fermion 
generations are localized towards the UV brane. Supersymmetry breaking causes the first two 
sparticle generations to decouple, thereby avoiding the supersymmetric flavour and $CP$ problems,
while an accidental $R$-symmetry protects the gaugino mass. 
The resulting low-energy sparticle spectrum consists of stops, gauginos and Higgsinos 
which are sufficient to stabilize the little hierarchy between the IR scale and the electroweak scale.
Finally, the supersymmetric little hierarchy problem is ameliorated by 
introducing a singlet Higgs field on the IR brane.

\end{abstract}

\vfill
\begin{flushleft}
\end{flushleft}
\eject
\pagestyle{empty}
\setcounter{page}{1}
\setcounter{footnote}{0}
\pagestyle{plain}

\section{Introduction} 
\label{secIntro}

\noindent
A warped extra dimension provides a natural way to explain hierarchies in the standard model.
In the five-dimensional ($5D$) warped geometry the local cutoff is position dependent. 
By localizing the Higgs sector on the infrared (IR) brane, the Higgs cutoff can be naturally
of order the TeV scale, thereby providing a solution to the gauge hierarchy problem~\cite{Randall:1999ee}. Furthermore by placing the standard model fermions in the bulk, large hierarchies in the Yukawa couplings can be explained by a wavefunction overlap with the Higgs boson in the extra dimension~\cite{Grossman:1999ra, Gherghetta:2000qt, Huber:2000ie}. This fermion and Higgs geography in the slice of AdS$_5$ therefore provides a novel framework to address hierarchies in the standard model without symmetries or hierarchies in the $5D$ Yukawa couplings (often referred to as anarchic couplings).

However, with anarchic couplings, $CP$-violating processes mediated by Kaluza-Klein (KK) modes are in excess of experimental bounds unless the IR scale is at least $\mathcal{O}(10 \text{ TeV})$ \cite{Agashe:2004cp,Bona:2007vi,Isidori:2010kg}. Although this bound can be avoided with additional structure (such as flavour symmetries, see e.g. \cite{flavoursymmetries}), electroweak precision tests still require an IR scale 
larger than the electroweak scale. To obtain the correct $Z$-boson mass, some tuning is needed. This is a manifestation of the little hierarchy problem \cite{Barbieri:2000gf} that also plagues other solutions to the gauge hierarchy problem.

A well-known way to protect the Higgs from radiative corrections is supersymmetry (SUSY). 
Usually, it is supposed to stabilize the entire hierarchy between the electroweak and the Planck scale. 
In this paper, we will instead entertain the possibility that SUSY protects the Higgs only up to 
$\mathcal{O}(10 \text{ TeV})$ and that warping (or compositeness in the dual picture) is responsible 
for the remaining hierarchy up to the Planck scale. For this purpose, a reduced form of SUSY is 
sufficient. Since the Higgs in warped models is localized near the IR brane, loops are cut off at a 
warped-down scale $\Lambda_\ir$. The one-loop correction to the Higgs mass due to a quark is\footnote{We will later consider the Higgs sector of the NMSSM, with two Higgs doublets and a singlet, and the (lightest $CP$-even) Higgs is an admixture of these fields. Depending on the mixing angles, up-type and/or down-type quarks couple additionally suppressed to this Higgs. Here and below, we will for simplicity assume that the admixture is such that Eq.~\eqref{one-loop} remains approximately valid.\label{HiggsCaveat}}
\be
\label{one-loop}
\Delta m^2_{H} \, = \, - \frac{3}{8 \pi^2} \, y_q^2 \, \Lambda_\ir^2 \, \sim - ( 10 \, m_q)^2,
\ee
where $y_q$ is the Yukawa coupling and $m_q$ the mass of the quark. In the last step, we have assumed that $\smash{\Lambda_\ir =\mathcal{O}(10 \text{ TeV})}$ and $\smash{\tan \beta = \mathcal{O}(1)}$.
In this case, only the top loop correction is in excess of the electroweak scale and stops are the only light superpartners required to protect the Higgs from the quark sector.\footnote{Then the left-handed sbottom is of course light as well. Sbottoms and staus become important for large $\tan \beta$. As we will discuss, however, we focus on small $\tan \beta$ in this paper.} Similarly, no lepton superpartners have to be light (or even present at all). Gauge bosons and the Higgs itself, on the other hand, lead to sizeable corrections whose cancellation requires light gauginos and Higgsinos.
This reduced spectrum of superpartners is all that is needed to protect the Higgs up to 
$\mathcal{O}(10 \text{ TeV})$. This is similar in spirit to Little Higgs models~\cite{ArkaniHamed:2001nc} except that our warped model provides a UV completion for energies above $10 \text{ TeV}$.

There is an important advantage if stops, gauginos and Higgsinos are the only superpartners near the electroweak scale. As is well-known, squarks and sleptons lead to excessive flavour and $CP$ violation 
if the mediation of SUSY breaking is not flavour-blind. These problems are avoided if the superpartners of the first two generations are very heavy. For this reason, highly non-degenerate sparticle spectra were considered already long ago \cite{Cohen:1996vb,MoreMinimal}. Naturalness, however, does not allow the relevant sparticles to become sufficiently heavy to entirely solve the flavour and $CP$ problems \cite{NaturalnessConstraints,ArkaniHamed:1997ab,Barbieri:2010pd1,Barbieri:2010pd2}. Since the hierarchy problem is mainly solved by warping/compositeness in our case, we have no such constraints. 

To decouple the superpartners of the first two generations is straightforward in a warped model \cite{Gherghetta:2003wm}. We break SUSY at a high scale on the ultraviolet (UV) brane. Since the light standard model fermions are localized near that brane, their superpartners feel SUSY breaking maximally and obtain high masses. Stops and Higgsinos, on the other hand, are localized near the 
IR brane and remain light. Generically, however, gauginos also obtain high masses since they have a sizeable wavefunction overlap with the UV brane. It was pointed out in \cite{Sundrum:2009gv} that the gauginos can be protected if the theory has an $R$-symmetry.\footnote{An alternative possibility to protect gauginos from SUSY breaking on the UV brane may be to localize the vector multiplets towards the IR brane. The localization of gauge bosons in warped space was discussed in \cite{gbl}.} This is analogous to how gauginos are kept light in split SUSY \cite{splitSUSY1,splitSUSY2}. 

\begin{figure}[t]
\centering 
\includegraphics[width=11cm]{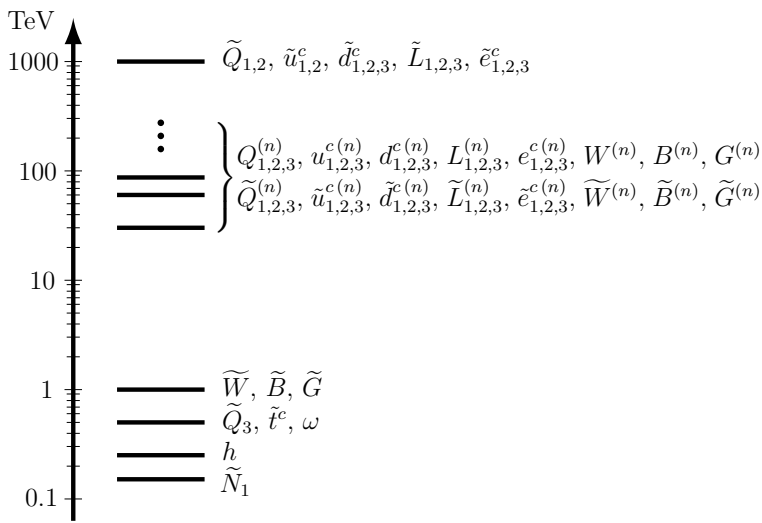}
\caption{Mass spectrum of superpartners and KK modes in our model: Gauge bosons and IR-localized particles
have superpartners near the electroweak scale, whereas scalar partners of UV-localized
fermions have masses above 1000 TeV. The KK towers of all particles start around 30 TeV. Note that the depicted spectrum below that scale is valid under certain (well-motivated) assumptions. These are discussed in Sect.~\ref{secHiggs}. The LSP $\smash{\widetilde{N}_1}$ is then dominantly Higgsino and the lightest $CP$-even Higgs $h$ is relatively heavy. We have not depicted the other fermions and scalars from the Higgs sector. The radion $\omega$ is stabilized around the soft scale. The gravitino, finally, is not shown as its mass is model-dependent. In the cases of interest, however, it is not the LSP.}
\label{Fig:mass.spectrum}
\end{figure}

In this paper, we extend the work of \cite{Gherghetta:2003wm} and \cite{Sundrum:2009gv} as follows:\footnote{The combination of SUSY and gauge-Higgs unification to stabilize the little hierarchy was considered in \cite{Redi:2010yv}.} Having protected gauginos, stops and Higgsinos from SUSY breaking on the UV brane, we have to transmit SUSY breaking to the IR brane. In order to generate a little hierarchy, the resulting soft masses of these superpartners should lie a factor $\mathcal{O}(100)$ below the IR scale (which is of order $\smash{10 \text{ TeV}}$). As the IR scale is set by the mechanism which stabilizes the extra dimension, we address the transmission of SUSY breaking and radion stabilization at the same time. To achieve both, we introduce a bulk hypermultiplet which obtains an $F$-term from the SUSY-breaking sector on the UV brane. We also add a constant superpotential on the IR brane. The energy density from both sectors leads to a radion-dependent potential and allows us to stabilize the extra dimension at an IR scale of order 10 TeV. Moreover, the bulk $F$-term naturally gives soft masses to superpartners in the IR which are a factor $\mathcal{O}(100)$ below the IR scale. Since these soft masses set the scale of electroweak symmetry breaking, our model generates and stabilizes the little hierarchy between the IR scale and the electroweak scale. The resulting mass spectrum is depicted in Fig.~\ref{Fig:mass.spectrum}.

Moreover, we emphasize the (well-known) fact that the Minimal Supersymmetric Standard Model (MSSM) has its own little hierarchy problem. 
It stems from the fact that the quartic Higgs coupling in the MSSM is determined by the electroweak gauge couplings. This results in an upper bound on the tree-level mass of the (lightest $CP$-even) Higgs,
which can never be larger than the $Z$-boson mass, in violation of the LEP bound. Although the Higgs mass can be raised by radiative corrections with heavy stops, this reintroduces some tuning to obtain the right vacuum expectation value (vev) for the Higgs. 

To avoid reintroducing the little hierarchy problem via SUSY, we include a singlet $S$ with a superpotential term $\lambda S H_u H_d$ (where $H_u$ and 
$H_d$ are the two Higgs superfields), analogous to the NMSSM. This gives an additional contribution to the Higgs quartic coupling that raises the Higgs mass already at tree-level.
If the NMSSM is the full theory to the GUT scale, the size of $\lambda$ is restricted by perturbativity making it difficult to raise the Higgs mass above the LEP bound; however,
in our warped model the singlet lives on the IR brane and $\lambda$ therefore has to be perturbative only up to the IR scale. 
Accordingly, the tree-level mass can be much larger than in the 
NMSSM. This framework, where the Higgs mass in SUSY is raised with large $\lambda$ but the theory is valid only up to $\smash{\mathcal{O}(10 \text{ TeV})}$, was dubbed $\lambda$SUSY in \cite{Barbieri:2006bg}. Our warped model can be viewed as a UV completion of $\lambda$SUSY.\footnote{The phenomenology of $\lambda$SUSY was further explored in \cite{Franceschini:2010qz,Kitano:2004zd}. Other UV completions were presented in \cite{Harnik:2003rs}.} Extending the Higgs sector has the further advantage that a vev of the singlet can generate the $\mu$-term at the right scale \cite{Franceschini:2010qz} (which would be too large if set by the IR scale).

As usual, our warped model can be given a dual interpretation: It corresponds to a strongly-coupled superconformal field theory which has weakly gauged global symmetries (corresponding to the standard model gauge group) and which couples to elementary fermions (leading to partially composite standard model fermions). SUSY is broken at a high scale in the UV (preserving an $R$-symmetry to protect the gauginos) but the theory runs towards a supersymmetric fixed point in the IR \cite{Sundrum:2009gv}. The hypermultiplet that obtains an $F$-term and a vev from the UV brane is dual to an operator that is added to the theory in the UV. This operator is responsible for the stabilization of the dilaton (dual to the radion) and the transmission of SUSY breaking to the IR. In this paper, however, we will focus on the gravity side of the gauge-gravity duality.

The outline of this paper is as follows: In Section 2 we show how the radion is stabilized by a SUSY-breaking potential on the UV brane and a constant superpotential on the IR brane.
This naturally generates an IR scale of order 10 TeV. In Section 3 we first calculate the effects of the SUSY breaking on the UV brane that is transmitted to the bulk standard model fields via a UV-localized messenger sector. The first-two-generation sparticles are shown to obtain masses above 1000 TeV, thereby alleviating flavour-violation problems, while gauginos, stops and Higgsinos remain light.  The transmission of SUSY breaking to the IR brane is then shown to give 
soft masses for stops and the Higgs sector which are hierarchically smaller than the IR scale. Gaugino masses of the same order are obtained either from radion mediation or from a suppressed operator on the UV brane. 
A discussion of the NMSSM Higgs sector on the IR brane and how it ameliorates 
the supersymmetric little hierarchy problem is given in Section 4. In Section 5 we provide further details of how the little hierarchy is generated. Electroweak and flavour constraints in our model are discussed in Section 6. In Section 7 we give concluding remarks. Finally, there are two appendices which provide further details on the stabilization mechanism and present an example messenger sector on the UV brane. 

\section{Radion stabilization}
\label{secRadionStabilization}

To construct models as outlined in the introduction, our starting point is a supersymmetric Randall-Sundrum model \cite{Gherghetta:2000qt,Altendorfer:2000rr}. 
The action for the supergravity multiplet is given in 
\cite{Gherghetta:2000qt,Altendorfer:2000rr} but will not be needed for our purposes. 
The background metric is \cite{Randall:1999ee}
\begin{equation}
ds^{2}=g_{MN}dx^M dx^N=e^{-2k |y|} \, \eta_{\mu\nu}dx^{\mu}dx^{\nu} + dy^2~,
\end{equation}
where $k$ is the AdS curvature scale and $\eta_{\mu\nu}={\rm diag}(-1,+1,+1,+1)$ is the Minkowski metric. The 5th dimension is an $S^1/\mathbb{Z}_2$ orbifold and $y$ denotes the 5th coordinate.
The orbifold fixed points, $y=0$ and $y=\ell$, are the positions of two 3-branes. 

Let us define and fix some scales for later use: As usual, we call the warped-down AdS scale $\smash{m_\ir \equiv e^{-k \ell} k}$ the IR scale. It determines the mass gap of KK modes $\smash{m_\kk \approx \pi m_\ir}$. We 
denote the cutoff (or the scale of new physics) on the IR brane by $\Lambda_\ir$ and the five- and four-dimensional Planck scales by $M_5$ and $M_4$, respectively. The former determines the latter via the relation $\smash{M_4^2 \simeq M_5^3/k}$. For simplicity, we will assume that $\smash{\Lambda_\ir = e^{-k \ell} M_5}$. To have at least one KK mode in each tower before the theory becomes strongly coupled, the cutoff $\Lambda_\ir$ has to be larger than the KK scale $m_\kk$. Since we want a relatively high IR scale, we need a little hierarchy $10 - 100$ between the IR scale and the electroweak scale $m_\ew$.
To stabilize this little hierarchy, the masses $\smash{m_\soft^\ir}$ of Higgsinos, gauginos and stops should lie only some small factor above the electroweak scale. For simplicity, we will often identify $\smash{m_\soft^\ir}$ with $m_\ew$ in numerical estimates.

We will now explain how we stabilize the extra dimension. 
For the stabilization, we follow \cite{Goh:2003yr} and include an additional hypermultiplet in the bulk. It is convenient to write this hypermultiplet in 
terms of two chiral superfields $\Phi$ and $\widetilde{\Phi}$ of $\smash{\mathcal{N}=1}$ SUSY. On both branes, we impose even boundary conditions on $\Phi$ and odd boundary conditions on $\widetilde{\Phi}$.
This breaks the $\mathcal{N}=2$ SUSY of the bulk theory down to $\smash{\mathcal{N}=1}$ SUSY on the branes. On the UV brane there is a sector which breaks the remaining $\mathcal{N} = 1$ SUSY completely. We assume that this sector induces an $F$-term for the hypermultiplet which we model with a SUSY-breaking potential $V$. 
The Lagrangian then reads \cite{Marti:2001iw}
\begin{multline}
\label{HypermLagr}
\mathcal{L}_5 \, \supset  \int d^4 \theta \, \left[ e^{-2 \sig} \left(\Phi^\dagger \Phi+ \widetilde{\Phi}^\dagger \widetilde{\Phi} \right) +  \, \delta(y) \, V(\Phi, F) \right]\\
\, + \, \left[\int d^2 \theta \, e^{-3 \sig} \, \widetilde{\Phi}  \left(   \partial_y  + \Bigl(c' - \frac{3}{2} \Bigr) k \epsilon(y)  \right) \Phi+ \text{h.c.} \, \right] 
 \,,
\end{multline}
where we parametrize the mass of the hypermultiplet by the dimensionless constant $c'$ and $\epsilon(y) \equiv \partial_y |y|$.
Due to the assignment of boundary conditions, $\widetilde{\Phi}$ does not couple to brane potentials and we have correspondingly 
omitted it from the SUSY-breaking potential $V$. We have written the SUSY-breaking potential as a superfield spurion, $V= - \theta^4 U$, with $U$ a function of the lowest component and $F$-term of $\Phi$. 
In an abuse of notation, 
we denote the lowest component by $\Phi$, whereas $F$ is the $F$-term. 
Similarly, the bosonic components of $\widetilde{\Phi}$ are respectively denoted by  $\widetilde{\Phi}$ and $\widetilde{F}$.

Expanding the hypermultiplet, $\smash{\Phi = \Phi + \theta^2 F + \dots}$ and $\smash{\widetilde{\Phi} = \widetilde{\Phi} + \theta^2 \widetilde{F} + \dots}$, the action yields the following equations
of motion for vevs of the scalars and $F$-terms:
\begin{gather}
\label{eom1}
e^{-3 \sig} \, \partial_y \Phi \,+  \, e^{-3 \sig} \Bigl(c'-\frac{3}{2}\Bigr) k \epsilon(y) \, \Phi \, + \, e^{-2 \sig}  \widetilde{F}^\dagger \, = \, 0 \\
\label{eom2}
\partial_y F \, + \, \Bigl(c'-\frac{3}{2}\Bigr) k \epsilon(y) \, F \, = \, 0\\
\label{eom3}
e^{-3 \sig} \, \partial_y \widetilde{\Phi} \,-  \, e^{-3 \sig} \Bigl(c'+\frac{3}{2}\Bigr) k \epsilon(y) \, \widetilde{\Phi} \, - \, e^{-2 \sig}  F^\dagger \, 
= \,- \delta(y) \frac{\partial U}{\partial F} \\
\label{eom4}
e^{-3 \sig} \, \partial_y \widetilde{F} \, - \, e^{-3 \sig} \Bigl(c'+\frac{3}{2}\Bigr) k \epsilon(y) \, \widetilde{F} \, = \, - \delta(y) \frac{\partial U}{\partial \Phi} \, .
\end{gather}
The bulk parts of these equations of motion are solved by
\begin{gather}
\label{bulksolutions1}
F \, = \, F_\uv \, e^{(4-\Delta) k|y|} \quad  \qquad \qquad \widetilde{F} \, = \,  \widetilde{F}_\ir  \, \rho^{\Delta-1} \, e^{(\Delta-1) k|y|}\epsilon(y)\\
\label{bulksolutions2}
\Phi\, = \, \left(\Phi_\uv \, + \, \frac{\widetilde{F}_\ir^\dagger \,  \rho^{\Delta-1} }{(2 \Delta-4)k} \right) \, e^{(4-\Delta) k|y|} \, - \, \frac{\widetilde{F}_\ir^\dagger \,  \rho^{\Delta-1} }{(2 \Delta-4)k} e^{\Delta k |y|} \\
\label{bulksolutions3}
\widetilde{\Phi} \, = \, \epsilon(y)\left[\left( \widetilde{\Phi}_\ir  \rho^{\Delta-1} \, + \, \frac{ F_\uv^\dagger \, \rho^{2 \Delta-6}}{(2 \Delta-6)k} \right) \, e^{(\Delta-1) k|y|} \, - \, 
\frac{F_\uv^\dagger \, }{(2 \Delta-6)k} e^{(5-\Delta) k |y|}\right] \, .
\end{gather}
We use the shorthands $\smash{\rho \equiv e^{-k \ell}}$ and $\smash{\Delta \equiv c' + \frac{5}{2}}$, where
$\Delta$ represents the dimension of the operator dual to the bulk field $\Phi$. As in \cite{Goh:2003yr}, we parametrize the bulk solutions by the values of 
$\Phi$ and $F$ at the UV brane and those of $\smash{\widetilde{\Phi}}$ and $\smash{\widetilde{F}}$ at the IR brane:
\be
\Phi_\uv \, \equiv \, \lim_{y \rightarrow 0^+} \Phi, \quad \quad \widetilde{\Phi}_\ir \, \equiv \, \lim_{y \rightarrow  \ell^-} \widetilde{\Phi} \quad \text{etc.}
\ee
These parameters are determined by boundary conditions which follow from the boundary pieces in the equations of motion:
\be
\begin{split}
\label{bc1}
\widetilde{\Phi}_\uv \, = \, -\frac{1}{2} \frac{\partial U}{\partial F_\uv} & \quad \; \quad \widetilde{F}_\uv \, = \, -\frac{1}{2} \frac{\partial U}{\partial \Phi_\uv} \\
\widetilde{\Phi}_\ir \, = \, 0 & \quad \; \quad \widetilde{F}_\ir \, = \, 0 \, .
\end{split}
\ee

Using the bulk wavefunctions and boundary conditions in Eq.~\eqref{HypermLagr} and integrating over the extra dimension, one obtains the contribution of 
the hypermultiplet to the effective four-dimensional ($4D$) potential. It will be useful to write this potential in a superfield form, for which one finds \cite{Goh:2003yr}
\be
\begin{split}
\label{superfieldform}
 V_4 \, \supset & \, \int d^4 \theta \left[- V + \frac{1}{2} \left( \Phi\frac{\partial V}{\partial \Phi} \, + \, F \frac{\partial V}{\partial F} \, + \,  \text{h.c.} \right)  \right]_\uv \\
  \, = & \, \left[ U(\Phi,F) \, + \, \left(\Phi\widetilde{F} \, + \, \widetilde{\Phi} F \, + \, \text{h.c.} \right) \right]_\uv \, ,
\end{split}
\ee
where the quantities in brackets are to be evaluated on the UV brane.

The potential $U(\Phi,F)$ explicitly breaks SUSY. We imagine that it originates from a supersymmetric coupling of the chiral multiplet $\Phi$ to a SUSY-breaking spurion on the UV brane. To fix the scales in the SUSY-breaking potential, we will assume that the relevant interactions are Yukawa couplings,
\be
\mathcal{L}_5 \, \supset \, \delta(y)  \int d^2\theta  \left(\frac{y_{ij}}{\sqrt{k}} \, \Phi Z_i Z_j \, + \, \frac{y_i}{k} \, \Phi^2 Z_i \right) \; + \; \text{h.c.} \, , 
\ee
where the chiral multiplets $Z_i$ are localized on the UV brane and couple to or belong to the SUSY-breaking sector (which contains the spurion). We have written the couplings in terms of the AdS scale and dimensionless constants $y_{ij}$ and $y_i$, whose nonvanishing entries we expect to be of order one. Under this assumption, each $\Phi$ and $F$ in the SUSY-breaking potential comes with a factor $\smash{k^{-1/2}}$. We will furthermore assume that the only other scale in the potential is the SUSY-breaking scale $M_\susyb$. As we will find below, we are interested in an intermediate scale $\smash{m_\ir \ll M_\susyb \ll M_5}$. This scale could arise from dynamical SUSY breaking or from SUSY breaking in the IR of another throat (both possibilities being related via gauge-gravity duality).

Let us now evaluate the effective $4D$ potential. For simplicity, we focus on a simple potential $U(\Phi,F)$ in the following. We show in Appendix \ref{App:Pot}, however, that our results on radion stabilization and SUSY breaking remain unchanged for a generic potential. More precisely, we consider a source term for $F$ and a mass term for $\Phi$:
\be
\label{SimpleU}
U(\Phi,F) \, = \, \left( e^{i\varphi_U} \,\frac{M_\susyb^2}{\sqrt{k}} \, F + \text{h.c.} \right) \, + \, \frac{M_\susyb^2}{k} \, |\Phi|^2 \, .
\ee
Here we have fixed the scales according to the assumption discussed before and $\varphi_U$ is a phase. Using the boundary conditions Eq.~\eqref{bc1}, we find:
\be
\begin{split}
\label{phi_uv}
\widetilde{F} \, \equiv \, 0 \qquad \quad & \Phi\, \equiv \, 0 \\
F_\uv^\dagger \, = \, e^{i \varphi_U} \,\frac{\Delta -3}{1- \rho^{2\Delta -6}} \sqrt{k} \, M_\susyb^2 & \qquad \widetilde{\Phi}_\uv \, = \, -\frac{e^{i \varphi_U}}{2} \frac{M_\susyb^2}{\sqrt{k}}\, .
\end{split}
\ee

Via $F_\uv$, the potential picks up a dependence on the radion $\smash{\rho = e^{-k \ell}}$. In a supersymmetric theory, the radion is part of a chiral multiplet \cite{Marti:2001iw} which has another 
scalar degree of freedom. To obtain the dependence of the potential on this additional scalar, we start from the superfield form of the potential in Eq.~\eqref{superfieldform} and promote the radion
$\rho$ to a chiral superfield $\omega$ by analytic continuation into superspace \cite{Luty:2000ec,Goh:2003yr}. As before, we denote the lowest component of the superfield $\omega$ by the same letter.
From Eq.~\eqref{superfieldform}, we then find the following contribution of the hypermultiplet to the $4D$ effective potential:
\be
\label{contribution1}
V_4 \, \supset \, \frac{1}{2} \, \frac{\Delta -3}{1- \omega^{2\Delta -6}}M_\susyb^4 \, + \, \text{h.c.} \, .
\ee

To stabilize the radion, we need another $\omega$-dependent contribution to the potential. 
To this end, we add a constant superpotential on the IR brane. As we will see shortly,
the energy density in the resulting minimum of the potential is positive. In order to tune the 
cosmological constant to zero, we therefore add another constant superpotential on the UV brane. Such constant superpotentials can for example arise from gaugino condensates. The resulting $4D$ effective Lagrangian is \cite{Luty:2000ec}
\be
\begin{split}
\label{SuperpLagr}
 \mathcal{L}_4 \, \supset & \, \frac{3 M_5^3}{k}  \int d^4 \theta \left(|\omega|^2 - |\phi|^2 \right) \, + \, \left[\int d^2 \theta \left(e^{i \varphi_\uv}  C_\uv^3 \, \phi^3 \, + \, e^{ i \varphi_\ir}  C_\ir^3 \, \omega^3 \right) \, + 
\, \text{h.c.}\right] \\
\supset & \,\frac{3 M_5^3}{k}  \left(|F_\omega|^2 - |F_\phi|^2 \right) \, + \, 3 \left( e^{ i \varphi_\uv} C_\uv^3 \, F_\phi \, + \, e^{i \varphi_\ir} C_\ir^3 \, F_\omega \,  \omega^2\, + 
\, \text{h.c.}\right)  \, ,
\end{split}
\ee
where $\smash{\phi = 1 + \theta^2 F_\phi}$ is the chiral compensator of supergravity. We have written the constant superpotentials in terms of mass scales $C_\uv$ and $C_\ir$ and phases $\varphi_\uv$ and $\varphi_\ir$, respectively. Note that the Lagrangian has no additional dependence on the $F$-terms $F_\phi$ and $F_\omega$ coming from Eq.~\eqref{superfieldform} since the $d^4 \theta$-integral has to act on the $\theta^4$ in the superfield spurion $V$ to give a nonzero result. The equations of motion are thus
\be
\label{F-terms}
 F_\phi^\dagger \, =  \,  e^{i \varphi_\uv} \,k \, \left(\frac{C_\uv}{M_5}\right)^3 \qquad \quad
 F_\omega^\dagger \, =  \, -  e^{i \varphi_\ir} \,k \, \left(\frac{C_\ir}{M_5}\right)^3 \, \omega^2\, .
\ee
Plugging the $F$-terms back into Eq.~\eqref{SuperpLagr} (and using the relation $\smash{M_4^2 \simeq M_5^3/k}$), the contributions of the constant superpotentials to the potential are
\be
\label{contribution2}
V_4 \, \supset \, \, 3 \, \frac{C_\ir^6}{M_4^2} \, |\omega|^4 \, -\,  3 \,  \frac{C_\uv^6}{M_4^2} \, .
\ee

We assume that the sector on the UV brane breaks SUSY at an intermediate scale $\smash{m_\ir \ll M_\susyb \ll M_5}$. We now use the hierarchy $M_\susyb/M_5$ to stabilize the radion $\omega$ at an exponentially small value. The 
$\omega$-dependent part of the potential Eqs.~\eqref{contribution1} and \eqref{contribution2} reads
\be
\label{PotExp}
V_{4} \, \supset \,  3 \, \frac{C_\ir^6}{M_4^2} \, |\omega|^4  \, + \,  ( \Delta -3) \, M_\susyb^4 \, |\omega|^{2\Delta -6}  \cos \gamma \, + \,  M_\susyb^4 \cdot \mathcal{O}\bigl( \omega^{4\Delta -12}\bigr) \, ,
\ee
where we have defined $\gamma \equiv (2\Delta - 6) \arg \omega$ and expanded the potential for $\Delta>3$ and $|\omega| \ll 1$.\footnote{More precisely, the expression for $\gamma$ is correct for $\smash{\arg \omega\in [-\pi,\pi]}$. For other values, $\arg \omega$  in $\gamma$ must be projected onto this interval via $2 \pi$-shifts. This ensures $2 \pi$-periodicity of $\smash{\cos \gamma}$ with respect to $\arg \omega$. } 
We first minimize with respect to $\arg \omega$. This fixes $\smash{\arg \omega= \pi/(2\Delta-6)}$, so that $\smash{\cos \gamma=-1}$.\footnote{More precisely, $\smash{\cos \gamma=-1}$ is the minimum for $\smash{\Delta\geq 3.5}$. 
Moreover, there are two such minima for $\smash{3.5<\Delta< 4.5}$ and even more minima for larger values of $\Delta$. For $\smash{3.25<\Delta<3.5}$, on the other hand, the minimum is at $\smash{0 > \cos \gamma > -1}$, whereas for $\smash{3<\Delta<3.25}$ there is no minimum with $\smash{\cos \gamma <0}$. Since we need negative $\cos \gamma$ to obtain the required minimum in the potential, we will exclude the latter case in the following.\label{GammaMinimum}} Minimizing with respect to $|\omega|$, we find that the
radion is stabilized at
\be
\label{RadionMinimum}
|\omega| \, \simeq \, \left[ \frac{\Delta-3}{\sqrt{6}} \, \frac{M_\susyb^2 \, M_4 }{C_\ir^3} \right]^{\frac{1}{5-\Delta}}  \, .
\ee

In Sect.~\ref{SUSYbrIR}, we will find that the little hierarchy between the electroweak scale (more precisely the soft scale on the IR brane) and the IR scale is determined by the constant superpotential on the IR brane:
\be
\label{littlehierarchy}
\frac{m_\soft^\ir}{m_\ir} \, \sim \, \left(\frac{C_\ir}{M_5}\right)^3 \, . 
\ee
A little hierarchy is thus already obtained for $C_\ir$ of order (but slightly smaller than) $M_5$. For $\smash{M_\susyb \ll M_5}$, we therefore have $\smash{ M_4M_\susyb^2\ll C_\ir^3}$ and a hierarchically small $|\omega|$ is obtained for $\Delta< 5$. A minimum with $\smash{\cos \gamma<0}$, on the other hand, is obtained for $\Delta>3.25$ (see footnote \ref{GammaMinimum}). We are therefore interested in the range $\smash{3.25<\Delta<5}$. Note that we can use Eqs.~\eqref{RadionMinimum} and \eqref{littlehierarchy} to express the SUSY-breaking scale $M_\susyb$ in terms of the hierarchy $\smash{m_\ir/k}$, the little hierarchy $\smash{m_\soft^\ir/m_\ir}$ and the parameter $\Delta$ (and the ratio $\smash{k/M_4}$). Once the hierarchy and the little hierarchy are fixed, either $M_\susyb$ or $\Delta$ remains as a free parameter of our model. We will constrain this parameter further in Sects.~\ref{SUSYbrRadion} and \ref{sec:GauginoMasses}.

We see from Eq.~\eqref{SuperpLagr} that the kinetic term of the scalar $|\omega|$ has a prefactor $M_5^3/k$, whereas the prefactor for the pseudoscalar $\arg \omega$ is $|\omega|^2 M_5^3/k$. After canonically normalizing these fields, we find that their masses in the potential minimum are
\be
m_\text{\tiny scalar} \, \sim \, m_\text{\tiny pseudoscalar} \, \sim \, \left( \frac{C_\ir}{M_5}\right)^3 \, m_\ir  \, \sim \, m_\soft^\ir \,.
\ee

We have so far implicitly assumed that the backreaction of $\widetilde{\Phi}$ and $F$ on the metric can be neglected. We should check whether this is justified. 
Since the Lagrangian and the energy-momentum tensor of the hypermultiplet can be expressed solely in terms of $\Phi$ and $\widetilde{\Phi}$, it is sufficient to perform this check for
$\widetilde{\Phi}$. Since $\smash{3.25<\Delta<5}$, we find that everywhere along the extra dimension
\be
\label{vevEstimate}
|\widetilde{\Phi}(y)|  \, < \, \frac{|F_\uv| \, }{(2 \Delta-6)k} \, e^{(5-\Delta) k |y|} \, \lesssim \, \frac{m_\soft^\ir}{m_\ir}\, M_5^{3/2} \,.
\ee
In the last step, we have evaluated the exponential on the IR brane and used Eqs.~\eqref{RadionMinimum} and \eqref{littlehierarchy}. With a little hierarchy between $\smash{m_\soft^\ir}$ and $\smash{m_\ir}$, the vev of $\smash{\widetilde{\Phi}}$ is thus everywhere much smaller than the Planck scale and the backreaction can be safely neglected.

Let us now show that the minimum Eq.~\eqref{RadionMinimum} is the global minimum. The $\omega$-dependent part of the potential to all orders in $|\omega|$ reads
\be
\label{allorderpotential}
V_{4} \, \supset\,  3 \, \frac{C_\ir^6}{M_4^2} \, |\omega|^4  \, + \, (\Delta-3) \, M_\susyb^4  \, \frac{1-|\omega|^{2 \Delta -6} \cos \gamma}{1 - 2 \, |\omega|^{2\Delta-6} \cos \gamma + |\omega|^{4 \Delta -12} } \, .
\ee
The only region, where the expansion leading to Eq.~\eqref{PotExp} is not valid, is for $\omega$ close to 1 (since the region $\omega >1$ is unphysical). The second  term in the potential is  
always positive and of order $\smash{ M_\susyb^4}$ for $\omega \leq 1$ and dominates in the minimum Eq.~\eqref{RadionMinimum}. In the region $\omega$ close to 1, on the other hand, the first term in the potential  dominates since $\smash{ M_4M_\susyb^2\ll C_\ir^3}$.  This shows that the minimum Eq.~\eqref{RadionMinimum} is indeed the global minimum.

As usual in supergravity, we have to tune a constant superpotential (on the UV brane in our case) in order to cancel the energy density coming from the SUSY-breaking sector. From Eqs.~\eqref{contribution2} and
\eqref{allorderpotential}, we see that we have to choose
\be
\label{cUV}
C_\uv^3 \, \simeq \, \sqrt{\frac{\Delta-3}{3}} \, M_4 \, M_\susyb^2 \, .
\ee
This fixes the mass of the gravitino
\be
m_{\psi_{3/2}} \, = \, \frac{C_\uv^3}{M_4^2} \, \sim \,  m_\soft^\ir \,\left(\frac{m_\ir}{k}\right)^{4-\Delta} \,,
\ee
where we have used Eqs.~\eqref{RadionMinimum} and \eqref{littlehierarchy}. The gravitino is thus the LSP when $\Delta<4$.

\section{Transmission of SUSY breaking}
\label{SUSYbrUV}
We will now discuss how standard model superpartners feel the SUSY breaking. We assume that quarks, leptons and gauge bosons live in the bulk and that the fermion mass hierarchy results from the localization of wavefunctions \cite{Grossman:1999ra,Gherghetta:2000qt}. 

We consider a generic abelian vector multiplet and hypermultiplet as representatives of the standard model multiplets in the bulk. 
It will be useful to write their action in terms of $\smash{\mathcal{N}=1}$ superfields. A bulk vector multiplet consists of
a vector multiplet $V$ and a chiral multiplet $\chi$ of $\smash{\mathcal{N}=1}$ SUSY and the Lagrangian reads \cite{Marti:2001iw}
\be
\label{vectormultiplet}
 \mathcal{L}_5\, \supset \, \left[ \frac{1}{4 } \int d^2 \theta \,  T \, W^\alpha W_\alpha \, + \, \text{h.c.}\right] \, + \, 2  \int d^4 \theta \, 
\frac{e^{-(T + T^\dagger)k |y|}}{T + T^\dagger} \left( \partial_y V - \frac{1}{\sqrt{2}}\left(\chi + \chi^\dagger \right)\right)^2 \, .
\ee
We impose even boundary conditions on $V$ and odd boundary conditions on $\chi$ at the two branes.  
For later use, we have included the dependence on the radion multiplet $T$. It is related to the radion multiplet $\omega$ of Sect.~\ref{secRadionStabilization} by
\be
\label{omegaphiT}
\omega \, = \, \phi \,  e^{-k \ell T}\, ,
\ee
where $\phi$ is the chiral compensator of supergravity. In our conventions, its lowest component has a vev $\smash{\langle |T| \rangle = 1}$ corresponding to $\smash{\langle |\omega| \rangle = e^{-k \ell}}$.

A hypermultiplet consists of two chiral multiplets $Q$ 
and $\widetilde{Q}$ of $\smash{\mathcal{N}=1}$ SUSY. The Lagrangian reads \cite{Marti:2001iw}
\begin{multline}
\label{hypermultiplet}
   \mathcal{L}_5\, \supset \, \int d^4 \theta \, \frac{1}{2} \left(T + T^\dagger\right) \, e^{-(T + T^\dagger) k |y|} \left(Q^\dagger   Q \, + \, \widetilde{Q}^\dagger \widetilde{Q} \right) \\
\, + \, \left[ \int d^2 \theta \, e^{-3 T k |y|} \, \widetilde{Q}  \left(   \partial_y  + \Bigl(c - \frac{3}{2} \Bigr) T  k\,  \epsilon(y)  \right) Q \, + \, \text{h.c.}\right]\, ,
\end{multline}
where we parametrize the mass of the hypermultiplet by the dimensionless constant $c$. We impose even boundary conditions on $Q$ and odd boundary conditions on $\widetilde{Q}$ at the two branes. For simplicity, we do not include couplings to vector multiplets.

The Higgs sector is localized on the IR brane. As motivated in the introduction, we consider the Higgs sector of the NMSSM, consisting of Higgs doublets $H_u$ and $H_d$ and a singlet $S$. 
The Lagrangian reads (see \cite{Luty:2000ec,Marti:2001iw} for the $\omega$-dependence)
\begin{multline}
\label{L5Higgs}
\mathcal{L}_5 \, \supset \, \delta(y - \ell) \, \int d^4 \theta \, \omega^\dagger \omega  \, \left(H_u^\dagger H_u \, + \, H_d^\dagger  H_d \,  + \, S^\dagger S \right)\\ + \,  \delta(y - \ell) \left[ \int d^2 \theta \, \omega^3 \left( \, y_u \, H_u Q Q  \, + \, y_d \, H_d Q Q \, + \,  \lambda \, S H_u H_d \, + \, \frac{\kappa}{3} \, S^3 \, \right) \, + \, \text{h.c.} \, \right] 
\, ,
\end{multline}
where $y_u$, $y_d$, $\lambda$ and $\kappa$ are Yukawa couplings and $Q$ represents both doublets and right-handed fields (respectively the corresponding bulk chiral multiplets). We have imposed a $\smash{\mathbb{Z}_3}$-symmetry under which $Q$ and the Higgs sector superfields\footnote{This $\mathbb{Z}_3$ is spontaneously broken  when the Higgses obtain vevs, leading to dangerous domain walls in the early universe \cite{Abel:1995wk}. Ways around this problem were e.g.~discussed in \cite{Abel:1996cr}.} transform with the phase $\smash{e^{2 \pi i/3}}$ whereas $\smash{\widetilde{Q}}$ transforms with the phase $\smash{e^{-2 \pi i/3}}$. This forbids dimensionful couplings in the superpotential and allows for a solution of the $\mu$-problem (see Sect.~\ref{secHiggs}).

\subsection{Contribution from UV-localized messengers}
\label{UVcontribution}
Heavy messengers transmit the SUSY breaking on the UV brane to the visible sector. We will assume that these messengers have Yukawa couplings to the standard model hypermultiplets and the spurion $\Phi$ and gauge couplings to the standard model vector multiplets.  We present a simple messenger sector along these lines in Appendix \ref{App:Mess}. For simplicity, we consider a messenger sector which is localized on the UV brane. Then due to the assignment of boundary conditions, only the chiral components $Q$ of standard model hypermultiplets and the components $V$ of standard model vector multiplets couple to the messengers $X_i$.  The relevant part of the Lagrangian reads
\begin{multline}
\label{MessengerSector}
 \mathcal{L}_5 \, \supset \, \delta(y)  \int d^2\theta  \left(\frac{y_1^{ij}}{\sqrt{k}} \, \Phi X_i X_j \, + \, \frac{y_2^i}{k} \, \Phi^2 X_i \,  + \, \frac{y_3^{ij}}{\sqrt{k}} \, Q X_i X_j \, + \, \frac{y_4^i}{k} \, Q Q X_i \right) \; + \; \text{h.c.} \, \\ + \,\delta(y) \int d^4\theta \, \left( g_5 \, X_i^\dagger  V  X_i  \, + \, \frac{g_5^2}{2} X_i^\dagger  V^2  X_i \right) \, ,
\end{multline}
where we have suppressed indices distinguishing the various standard model multiplets $Q$ and $V$.
We have written the Yukawa couplings in terms of the AdS scale (instead of e.g.~the $5D$ Planck scale) and dimensionless matrices $\smash{y_1^{ij}}$ to $\smash{y_4^i}$. We assume that the entries of these matrices are of order one (if they are nonzero). The gauge couplings $g_5$ furthermore are of order $\smash{k^{-1/2}}$. Integrating out the messengers, in particular we obtain the term
\be
\label{UVlc1}
\mathcal{L}_5 \, \supset \, \delta(y) \, \int d^4 \theta \, \frac{\Phi^\dagger \Phi}{k^2 M_X^2} \,  Q^\dagger Q \, ,
\ee
where $M_X$ is the messenger mass scale. Inserting the wavefunction of the scalar zero-mode (see e.g.~\cite{Gherghetta:2000qt}) of the chiral multiplet $Q$, we find the soft mass
\begin{equation}
m_{\tilde{q}}^\uv
	\sim 	\frac{|F_\uv|}{\sqrt{k} M_X } \sqrt{\frac{\frac{1}{2}-c}{e^{2 k \ell (\frac{1}{2}-c)}-1}} 
	\sim 	\frac{M_\susyb^2}{M_X }  \times \begin{cases} 
							e^{-k \ell (\frac{1}{2}-c)}	& c < \frac{1}{2} \quad \text{(IR-localized)}		\\
							1				& c > \frac{1}{2} \quad \text{(UV-localized)}
						\end{cases}
\label{scalarmasses}
\end{equation}
for this particle. The last step is valid for $\smash{\frac{1}{2}-c=\mathcal{O}(1)}$.

The messenger sector is generically not flavour-blind and we therefore have to have masses $m_{\tilde{q}} \gtrsim 1000$ TeV for the superpartners
of the first and second generation to avoid excessive flavour and $CP$ violation (see \cite{Barbieri:2010pd1} for a recent analysis). At the same time we have to keep gauginos and stops light. To protect the gauginos, we 
charge the stabilizing hypermultiplet (and thus the SUSY-breaking spurion $\Phi$) under a $\smash{U(1)'}$ gauge symmetry.
This forbids the coupling $\smash{\Phi W_\alpha W^\alpha}$ and the leading contribution to gaugino masses generically is\footnote{In Sect.~\ref{sec:R-symmetry} we discuss scenarios in which this term does not arise.} 
\be
\label{UVlc2}
\mathcal{L}_5 \, \supset \, \delta(y) \, \int d^4 \theta \, \frac{\Phi^\dagger \Phi}{k^2 M_X^3 } \, W^\alpha W_\alpha \, + \, \text{h.c.}\, .
\ee
Inserting the wavefunction of the gaugino zero-mode (which is of order $\ell^{-1/2}$ on the UV brane, see e.g.~\cite{Gherghetta:2000qt}), we find the soft mass of this particle
\be
\label{gauginomass1}
m_{\tilde{g}}^\uv \, \sim \, \frac{M_\susyb^4}{ k\ell  \, M_X^3}\, .
\ee
This is suppressed by a factor $\smash{M_\susyb^2/(M_X^2 k \ell)}$ compared to the soft mass of UV-localized scalars. 
We have to ensure that this contribution to gaugino masses is not larger than the soft scale on the IR brane. This gives a lower bound on the messenger scale
\be
\label{MXbound}
M_X \, \gtrsim \, \frac{M_\susyb^{4/3}}{m_\ir^{1/3}} \, ,
\ee
where we have assumed that $\smash{m_\ir \sim k \ell \, m_\soft^\ir}$ (with $\smash{k \ell \approx 30}$). Using this result in Eq.~\eqref{scalarmasses}, we find that the scalar masses for the first two generations (which have $\smash{c > \frac{1}{2}}$) can become as large as
\be
\label{smb}
m_{\tilde{q}}^\uv \, \lesssim \, m_\ir^{1/3} \, M_\susyb^{2/3}  \, \lesssim \, 10^{9} \text{ GeV}\,.
\ee 
In the last step, we have used a result that is derived in Sect.~\ref{SUSYbrRadion}: To avoid too large gaugino masses from radion mediation, the SUSY-breaking scale is constrained by $\smash{M_\susyb \lesssim \sqrt{m_\ir \, M_4}}$. We have furthermore assumed that $\smash{m_\ir \approx 10 \text{ TeV}}$. The soft masses for the first two generations can thus be much larger than the 1000 TeV  required to avoid problems with flavour and $CP$ violation if the messenger sector is not flavour-blind.

But we have to ensure that the stops do not pick up too large masses from the UV brane. These masses strongly depend on the parameter $c$ (see Eq.~\eqref{scalarmasses}). In Ref.~\cite{Huber:2003tu}, assuming an IR scale of 3 TeV, a statistical analysis was used to find that the `most natural' values for the top-bottom doublet and the right-handed top are $c=0.317$ and $c=-0.460$, respectively. With the former $c$-value, the ratio between the soft masses of the left-handed stop (the first line in Eq.~\eqref{scalarmasses}) and the first-two-generation sparticles (the second line in Eq.~\eqref{scalarmasses}) is of order $\smash{10^{-3}}$. To ensure that the former is lighter than a TeV, the latter can therefore at most be 1000 TeV. We note, however, that small changes in the $c$-values of IR-localized fields can easily be compensated by small changes in the $5D$ Yukawa couplings because their wavefunction overlap with the IR brane goes only like $\smash{\sqrt{1/2-c}}$. This allows to localize the top-bottom doublet more towards the IR  brane which in turn allows for larger soft masses on the UV brane.\footnote{We have to be careful with localizing the top-bottom doublet more towards the IR brane, though, as it tightens constraints from the $Z\bar{b}b$-coupling. Our IR scale is much higher than usual, on the other hand, which eases these constraints. The viability of this possibility also depends on whether we impose a custodial symmetry or not (see Sect.~\ref{sec:Constraints}).} Alternatively, if we want to keep the $5D$ Yukawa couplings fixed, we can move all left-handed fields (and in particular the top-bottom doublet)  towards the IR brane so that their wavefunction overlap with that brane is changed by a common factor. To compensate for this, we can then move the right-handed fields away from the IR brane (see \cite{Huber:2003tu}).

\subsection{Breaking of the $\smash{U(1)'}$}
\label{U(1)'breaking}

The $\smash{U(1)'}$ is broken by the vev of $\smash{\widetilde{\Phi}}$ (see Eq.~\eqref{phi_uv}) and the $\smash{U(1)'}$ gauge boson obtains a bulk mass $\smash{g'_5\widetilde{\Phi}}$. This lifts the massless mode of the gauge boson. 
To estimate its mass, we insert the constant wavefunction of the formerly massless mode into the bulk mass term and integrate over the extra dimension. Choosing $\smash{g_5'=\mathcal{O}(k^{-1/2})}$, this gives the estimate
\be
m_\text{zero-mode} \; \sim \;  \, \frac{m_\soft^\ir}{\sqrt{k \ell}}\,\frac{M_4}{k} \, \Bigl[ \, 1 \,  + \, \Bigl(\frac{k}{m_\ir}\Bigr)^{\Delta-4} \Bigr] \, .
\ee
This mass lies around or above the soft scale $\smash{m_\soft^\ir}$. The standard model is not charged under $\smash{U(1)'}$ and it therefore couples to the gauge boson only via $\smash{\Lambda_\ir^{-1}}$-suppressed operators. A mass of order the soft scale is then sufficient to avoid collider constraints.

Note that the $\smash{U(1)'}$ is explicitly broken by the simple SUSY-breaking potential in Eq.~\eqref{SimpleU}. We emphasize, though, that this potential is just an example and that other potentials work equally well as shown in Appendix \ref{App:Pot}. It is furthermore straightforward to check that this also applies to $\smash{U(1)'}$-invariant potentials. Since $\smash{\widetilde{\Phi}}$ (and generically $\Phi$ as well) obtain vevs, these potentials  break the $\smash{U(1)'}$ spontaneously. 

These vevs are $\smash{U(1)'}$-breaking spurions that are necessarily present in our model. We have to check whether these spurions lead to additional contributions to gaugino masses. As the combination $\smash{\Phi \widetilde{\Phi}}$ is gauge invariant (see Eq.~\eqref{HypermLagr}), the vev of $\smash{\widetilde{\Phi}}$ allows for gaugino masses from the coupling
\be
\label{PhiTildegm}
\delta(y) \int  d^2 \theta \; \Phi \partial_y \widetilde{\Phi} \, W_\alpha W^\alpha \,+ \,  \text{h.c.} \, .
\ee
Here we have taken into account that $\smash{\widetilde{\Phi}}$ (being odd) couples only derivatively to the branes.
This term is for example generated if $\smash{\partial_y \widetilde{\Phi}}$ couples to the messengers, in analogy to Eq.~\eqref{MessengerSector}. Assuming that
the coupling strength is of order $\smash{k^{-3/2}}$, the term in Eq.~\eqref{PhiTildegm} is suppressed by
\be
\frac{1}{k^3M_X^2}\, .
\ee
Using the results for $F$ and $\smash{\widetilde{\Phi}}$ from Sect.~\ref{secRadionStabilization}, we find that the resulting gaugino mass is suppressed by a factor $\smash{M_X/k}$ compared to Eq.~\eqref{gauginomass1}. 
The vev of $\Phi$ (see Appendix \ref{App:Pot}), on the other hand, does not lead to additional contributions to gaugino masses:  By SUSY and gauge invariance, the leading coupling of $\Phi$ to the gauge field strength is
\be
\delta(y) \int d^4 \theta \, \Phi^\dagger \Phi \,  W_\alpha W^\alpha \,+ \,  \text{h.c.} \, .
\ee
This is the coupling already considered in Eq.~\eqref{UVlc2} and the resulting gaugino mass does not depend on the vev of $\Phi$. The only terms that depend on this vev after performing the $\smash{d^4 \theta}$-integral  are small contributions (suppressed by a factor $\smash{(M_\susyb/M_X)^3}$) to the kinetic terms of the gauge multiplet.

\subsection{Contribution from gravity mediation in the IR}
\label{SUSYbrIR}
In addition to the soft masses discussed in the last section, which are induced by messengers on the UV brane, contributions to soft masses also arise from gravity mediation in the bulk (and on the branes).

We will assume that the third-generation quark doublet and the right-handed top are sufficiently IR-localized that the gravity-mediated contributions dominate. 
Since we contend ourselves with $\mathcal{O}(1)$-precision, we can replace these IR-localized bulk fields by brane-localized fields. Their $4D$ effective Lagrangian then reads (see \cite{Luty:2000ec} for the $\omega$-dependence)
\be
\label{L4eff}
\mathcal{L}_4 \, \supset \,  \int d^4 \theta \, \omega^\dagger \omega  \,Q^\dagger Q \, + \, \left( \int d^2 \theta \, \omega^3 \, H_u Q Q  \, + \, \text{h.c.}\right)  \,,
\ee
where $Q$ represents both the third-generation quark doublet and the right-handed top. The IR-localized sector also contains the Higgs multiplets whose Lagrangian is given in Eq.~\eqref{L5Higgs}.
Soft masses for IR-localized multiplets arise from the gravity-mediated couplings \cite{Luty:2000ec,Goh:2003yr}
\be
\mathcal{L}_4 \, \supset \,  \int d^4 \theta \, \omega^\dagger \omega \, \frac{\left[\Phi^\dagger \Phi\right]_\ir}{M_5^3}  \left(Q^\dagger Q  + H_u^\dagger H_u + H_d^\dagger H_d  + S^\dagger S\right) 
\label{PhiCouplingIR}
\ee
in the $4D$ effective Lagrangian, where the bulk spurion $\Phi$ is evaluated on the IR brane. The radion superfield $\omega$ determines the appearance of the warp factor in the Lagrangian. Since dimensionful couplings are forbidden by the $\mathbb{Z}_3$-symmetry, it can be completely eliminated from the Lagrangian in the last two equations and Eq.~\eqref{L5Higgs} via the field redefinitions $\omega Q \rightarrow Q$ etc. Expanding in $\theta$ in Eq.~\eqref{PhiCouplingIR}, we find the soft masses
\be
\label{msoftIR}
m_{\soft}^\ir \, = \, \frac{|F_\ir|}{M_5^{3/2}} \, \simeq \,  (\Delta-3)\,  \frac{M_\susyb^2}{M_4} |\omega|^{\Delta-4} \, \simeq \, \sqrt{6} \, \left(\frac{C_\ir}{M_5}\right)^3 \, m_\ir \,.
\ee
Here we have used Eqs.~\eqref{bulksolutions1} and \eqref{RadionMinimum}. We thus find that a little hierarchy between the soft scale and the IR scale arises when the ratio $\smash{(C_\ir/M_5)^3}$ is small. Due to the third power involved, a little hierarchy is in turn already obtained from a very modest hierarchy between $C_\ir$ and $M_5$. We will discuss this little hierarchy in more detail in Sect.~\ref{sec:LittleHierarchy}.

The couplings $\smash{\Phi W_\alpha W^\alpha}$ and $\smash{\Phi W_\ir}$ would generate gaugino masses and  $A$-terms around the soft scale $m_\soft^\ir$. However, these couplings are forbidden by the $\smash{U(1)'}$ and these soft terms are therefore suppressed:
\begin{equation}
\begin{split}
 \label{R-violating-1}
\mathcal{L}_5 \, \supset \,  \delta(y-  \ell) \int d^4 \theta  \,\frac{\omega^\dagger}{\omega^2} \, \frac{\Phi^\dagger \Phi}{M_5^5} \, W_\alpha W^\alpha \, + \, \text{h.c.} \quad \Rightarrow & \quad \smash{m_{\tilde{g}}^\ir \, \sim \, \frac{\left(m^\ir_\soft\right)^2}{k \ell \, m_\ir}\, }\\
\mathcal{L}_5 \, \supset \,  \delta(y-  \ell)  \int d^4 \theta \, \omega^\dagger \omega \, \frac{\Phi^\dagger \Phi}{M_5^4} \, W_\ir \, + \, \text{h.c.} \quad \Rightarrow & \quad \smash{A^\ir \, \sim \,\frac{\left(m^\ir_\soft\right)^2}{m_\ir}\, } \, .
\end{split}
\end{equation}
In the last step, we have ignored the numerical factor between $k$ and $M_5$. We will discuss ways to obtain gaugino masses and $A$-terms of $\smash{\mathcal{O}(m_\soft^\ir)}$ in Sect.~\ref{sec:GauginoMasses}. Note that we have assumed that the coupling $\smash{\Phi^\dagger \Phi W_\alpha W^\alpha}$ exists only on the branes (the gravity-mediated coupling on the UV brane is discussed in the next section). If this coupling is also present in the bulk, the gaugino mass can be enhanced by a factor $k \ell$ depending on the profile of $F$. Note furthermore that the vev of $\smash{\widetilde{\Phi}}$ (see Sect.~\ref{U(1)'breaking}) allows for additional contributions to gaugino masses and $A$-terms. These are however of the same size as those in Eq.~\eqref{R-violating-1}.  No additional contributions arise, on the other hand, from a vev of $\Phi$: The only terms that depend on this vev after performing the $\smash{d^4 \theta}$-integrals in Eq.~\eqref{R-violating-1}  are small contributions (suppressed by a factor $\smash{(m^\ir_\soft)^2/(M_\susyb m_\ir)}$) to gauge kinetic terms and superpotential terms.

\subsection{An accidental $R$-symmetry}
\label{sec:R-symmetry}
We have found that gaugino masses and $A$-terms are suppressed compared to soft scalar masses. This suppression can be understood as follows: 
The visible sector de-scribed by the Lagrangian in Eqs.~\eqref{vectormultiplet}--\eqref{L5Higgs} has an $R$-symmetry if we assign the charges
\be
R[V] \, = \, R[\chi] \, = \, 0 \, , \quad R[Q] \, = \, R[H_u] \, = \, R[H_d] \, = \, R[S] \, = \, \frac{2}{3} \quad \text{and} \quad  R[\widetilde{Q}] \, = \, \frac{4}{3} \,.
\ee
This symmetry is accidental because it was not imposed on the theory. The SUSY-breaking spurion appears only in the $\smash{U(1)'}$-invariant combination 
\be
\label{spurion}
\Phi^\dagger \Phi \, = \, \theta^4 \, |F|^2\, ,
\ee
where we have used that $\smash{\Phi \equiv 0+\theta^2 F}$.\footnote{For generic SUSY-breaking potentials, $\Phi$ usually obtains a vev (see Appendix \ref{App:Pot}). The vev of $\smash{\widetilde{\Phi}}$ moreover allows for another gauge-invariant SUSY-breaking spurion. We have seen in Sects.~\ref{U(1)'breaking} and \ref{SUSYbrIR}, however, that the resulting contributions to gaugino masses and $A$-terms are comparable or subdominant to those due to the spurion in Eq.~\eqref{spurion}. We will therefore ignore these vevs in the following discussion.} Since $\smash{|F|^2}$ has vanishing $R$-charge, this effective spurion does not break the accidental $R$-symmetry.\footnote{As the spurion has only a $\smash{\theta^4}$-component, this case was dubbed `$D$-breaking' in \cite{splitSUSY2}.} This symmetry prevents gaugino masses and $A$-terms at leading order in the messenger scale. The messenger sector on the UV brane, however, generically breaks the $R$-symmetry and generates the $R$-violating term in Eq.~\eqref{UVlc2} at higher order in $\smash{M_X^{-1}}$. Similarly, we do not expect that gravity respects the $R$-symmetry.\footnote{If gravity preserves the $R$-symmetry, the leading contribution to gaugino masses and $A$-terms involves an insertion of the constant superpotentials (which explicitly break the $R$-symmetry)  \cite{splitSUSY2}.} Gravity loops then induce the $R$-violating terms in Eq.~\eqref{R-violating-1} at higher order in $\smash{M_5^{-1}}$. This additional suppression of $R$-violating operators by the respective messenger scale is the reason for the smallness of gaugino masses and $A$-terms. 

This argument shows how the contribution to gaugino masses from the UV brane can be even further suppressed: If the messengers respect the accidental $R$-symmetry, the term in Eq.~\eqref{UVlc2} is not generated. We present a simple messenger sector with this feature in Appendix \ref{App:Mess}. In this case, the leading contribution to gaugino masses on the UV brane is mediated by gravity:
\be
\label{gmgmUV}
\mathcal{L}_5 \, \supset \, \delta(y) \, \int d^4 \theta \,  \frac{\Phi^\dagger \Phi}{M_5^5 } \, W^\alpha W_\alpha \, + \, \text{h.c.}\quad \Rightarrow  \quad m_{\tilde{g}}^\uv \, \sim \, \frac{M_\susyb^4}{k \ell \, M_5^3}\, . 
\ee
In the last step, we have ignored the numerical factor between $k$ and $M_5$. This term is analogous to the gravity-mediated gaugino mass term on the IR brane in Eq.~\eqref{R-violating-1}.
Since it does not depend on the messenger scale $M_X$, no bound on that scale (as in Eq.~\eqref{MXbound}) arises from the requirement that gauginos are sufficiently light. Correspondingly, soft scalar masses can be raised almost to the SUSY-breaking scale:
\be
m_{\tilde{q}}^\uv \, \lesssim \, M_\susyb \,.
\ee
The bound on the SUSY-breaking scale coming from Eq.~\eqref{gmgmUV} is in turn much weaker than another bound that we will derive in the next section.

\subsection{Contribution from radion mediation}
\label{SUSYbrRadion}

We will now discuss an additional source of $R$-symmetry breaking: From Eq.~\eqref{F-terms} (and since $\smash{\omega = \phi e^{-k\ell T}}$), we see that the radion superfield $T$ obtains an $F$-term
\be
F_T \, = \, \frac{1}{k \ell} \left(F_\phi - \frac{F_\omega}{\omega}\right) \, \sim \, \frac{C_\uv^3  +\omega \,  C_\ir^3 }{M_5^3 \ell}\, .
\ee
Assigning $R[T]=0$ to make the Lagrangian Eq.~\eqref{vectormultiplet} $R$-invariant, we see that the $F$-term breaks the accidental $R$-symmetry of Sect.~\ref{sec:R-symmetry}.
This corresponds to the fact that the $F$-term induces a bulk gaugino mass $\sim F_T$ in the Lagrangian Eq.~\eqref{vectormultiplet}. Assuming that $F_T \ll m_\ir$, only the zero-mode is significantly affected by this bulk mass. Inserting its wavefunction in Eq.~\eqref{vectormultiplet} and integrating over the extra dimension, one finds that the zero-mode obtains the mass  \cite{Chacko:2000fn,Marti:2001iw}
\be
\label{gauginomass2}
m_{\tilde{g}}^T \, = \, \frac{F_T}{2} \, \sim \, \frac{m_\soft^\ir}{k \ell} \,  \left[1 \, + \, \Bigl(\frac{m_\ir}{k}\Bigr)^{4-\Delta} \right]  \, .
\ee
In the last step, we have expressed $C_\uv$ and $C_\ir$ in terms of the hierarchy $\smash{m_\ir/k}$, the little hierarchy $\smash{m_\soft^\ir/m_\ir}$ and the parameter $\Delta$ using the results of Sect.~\ref{secRadionStabilization}. To ensure that gauginos remain light, we need $F_T \ll m_\ir$ so that the initial assumption is fulfilled. More precisely, the requirement that $\smash{m_{\tilde{g}}^T \lesssim m_\soft^\ir}$ restricts the parameter $\Delta$ to the range
\be
\label{DeltaCondition}
\Delta \, \leq \, 4 \, + \, \delta \quad  \quad \text{where} \quad \delta \, \simeq \,  \frac{\ln(k \ell)}{k \ell} \, .
\ee
Using Eqs.~\eqref{RadionMinimum} and \eqref{littlehierarchy}, this can equivalently be written as a condition on the SUSY-breaking scale $M_\susyb$. 
Assuming that $\smash{m_\ir \sim  m_\soft^\ir k \ell}$ (with $\smash{k \ell \approx 30}$), this gives
\be
\label{GauginoCondition}
M_\susyb \, \lesssim \, \sqrt{m_\ir \, M_4} \, .
\ee

Finally, let us discuss radion-mediated contributions to other soft terms. The radion $F$-term induces a soft scalar mass in the hypermultiplet Lagrangian Eq.~\eqref{hypermultiplet}. For $\smash{F_T \ll m_\ir}$, the zero-mode obtains the mass \cite{Marti:2001iw}
\be
\label{tree-level-FT}
m_{\tilde{q}}^T \, = \, \left|\frac{(\frac{1}{2}- c) \, k \ell \, F_T}{2 \sinh\left[(\frac{1}{2}-c)  k \ell \right]} \right| \, .
\ee
This contribution is maximal, $\smash{m_{\tilde{q}}^T = F_T/2 \lesssim m_\soft^\ir}$, for hypermultiplets with $\smash{c=1/2}$ but quickly becomes smaller away from $\smash{c=1/2}$. Since hypermultiplets with $c$ close to $1/2$ obtain much larger soft masses from the UV brane (see Eq.~\eqref{scalarmasses}), it is negligible.

On the IR brane, the radion superfield $\omega$ (which contains $T$) can be eliminated from the Lagrangian via field redefinitions (see Sect.~\ref{SUSYbrIR}).\footnote{Correspondingly, the tree-level contribution Eq.~\eqref{tree-level-FT} vanishes for IR-localized superfields in the limit $c\rightarrow - \infty$.} 
But soft terms sourced by $F_T$ nevertheless arise via anomaly mediation. The order parameter for anomaly mediation on the IR brane is $F_\omega/\omega$ \cite{Luty:2002ff} so that
\be
m_\soft^\amsb \, \sim \, \frac{F_\omega}{16 \pi^2\,  \omega} \, \sim \, \frac{m_{\soft}^\ir}{16 \pi^2} \, \ll \, m_\soft^\ir \, .
\ee
In the last step, we have used Eqs.~\eqref{F-terms} and \eqref{msoftIR}. This contribution is again negligible.

\subsection{Gaugino masses}
\label{sec:GauginoMasses}

To generate gaugino masses and $A$-terms of $\smash{\mathcal{O}(m_\soft^\ir)}$, we could break the $\smash{U(1)'}$ again (cf.~Sect.~\ref{U(1)'breaking}) on the IR brane to allow the couplings $\smash{\Phi W_\alpha W^\alpha}$ and $\smash{\Phi W_\ir}$. Once these terms are allowed, however, a tadpole for $\Phi$ is also allowed:
\be
\mathcal{L}_5 \, \supset \, \delta(y- \ell)  \int d^2\theta \, \omega^3 M_\text{0}^{3/2} \, \Phi \, + \, \text{h.c.} \,,
\ee
where $M_\text{0}$ is a mass scale. This superpotential term gives a correction of order \cite{Goh:2003yr}
\be
\sqrt{k} \, M_\text{0}^{3/2}  M_\susyb^2 \, \omega^{\Delta-1} \, + \, \text{h.c.}
\ee
to the potential Eq.~\eqref{PotExp}. By comparing the correction evaluated in the minimum Eq.~\eqref{RadionMinimum} with the other terms in the potential, we see that for 
\be
\label{DestabCon}
M_\text{0} \, > \, M_5 \, \left(\frac{m_\soft^\ir}{m_\ir}\right)^{2/3}
\ee
the minimum is destabilized. Since natural scales for $M_\text{0}$ are $M_5$ or $k$, we expect that this generically happens when we break the $\smash{U(1)'}$. 
We therefore choose not to break the $\smash{U(1)'}$ additionally on the IR brane.\footnote{Note moreover that even if the coupling $\smash{\Phi W_\alpha W^\alpha}$ is allowed on the IR brane, the resulting gaugino mass is volume-suppressed by the factor $k \ell$ compared to the soft scale $\smash{m^\ir_\soft}$.}

To generate sufficiently large gaugino masses, we could use the radion-mediated contribution discussed in the last section. Gaugino masses of $\mathcal{O}(m_\soft^\ir)$ are obtained when the bounds in Eqs.~\eqref{DeltaCondition} and \eqref{GauginoCondition} are saturated. To this end, we have to choose
\be
\Delta \, \simeq \, 4.1 \quad \quad \text{and} \quad \quad M_\susyb \, \sim \, 10^{11}  \text{ GeV}
\ee
for $\smash{m_\ir = \mathcal{O}(10\text{ TeV})}$. This SUSY-breaking scale is comparable to the intermediate scale considered in models of anomaly-mediated or gravity-mediated SUSY breaking. 
The parameter $\Delta$, on the other hand, is the dimension of the operator dual to the multiplet $\Phi$. This dual operator is thus marginally irrelevant.

However, Eq.~\eqref{gauginomass2} shows that radion-mediated gaugino masses and soft masses of IR-localized fields are of the same order only for finely tuned $\Delta$ due to the exponential dependence on $\Delta$. To avoid this issue, 
we make use of the messenger-mediated contribution to gaugino masses in Eq.~\eqref{gauginomass1}. This has the advantage that the gaugino masses have only a power-law dependence on the messenger scale $M_X$ which therefore does not have to be finely tuned. Gaugino masses of $\smash{\mathcal{O}(m_\soft^\ir)}$ are obtained if we choose the messenger scale
(assuming $\smash{m_\ir \sim k \ell \, m_\soft^\ir}$)
\be 
M_X \, \sim \, \frac{M_\susyb^{4/3}}{m_\ir^{1/3}} \, . 
\label{M_XforUVop} 
\ee 
To ensure that gauginos obtain their masses dominantly from messenger mediation and not from radion mediation, we have to require that $\smash{\Delta}$ satisfies the bound in Eq.~\eqref{DeltaCondition} or equivalently that $\smash{M_\susyb \, \lesssim \, 10^{11} \text{ GeV}}$. To have masses $m_{\tilde{q}} \gtrsim 1000 \text{ TeV}$ for the superpartners of the first and second generation (thereby avoiding the SUSY flavour problem), the messenger scale has to fulfill 
\be 
M_X \, \lesssim \, \frac{M_\susyb^2}{\text{1000 TeV}}\, , 
\ee 
where we have used eq.~\eqref{scalarmasses}. Assuming $\smash{m_\ir = \mathcal{O}(\text{10 TeV})}$ in eq.~\eqref{M_XforUVop}, we then see that the SUSY-breaking scale has to satisfy $\smash{M_\susyb \gtrsim 10^7 \text{ GeV}}$ to obtain sufficiently large scalar masses for the first two generations. These two requirements restrict the range of the remaining free parameter, $M_\susyb$ or equivalently $\Delta$, of our model.

\section{Electroweak symmetry breaking and the LSP}
\label{secHiggs}
We will now review aspects of electroweak symmetry breaking and the mass spectrum in models with large Higgs-singlet coupling $\lambda$.
The Higgs potential is determined by the superpotential in Eq.~\eqref{L5Higgs} and the $D$-term contributions
\be
V_\text{Higgs} \, \supset \, \frac{1}{8} \left(g_1^2 + g_2^2 \right) \, \left(|H_u|^2 - |H_d|^2 \right)^2 \, + \, \frac{1}{2} \, g_2^2 \left|H_d^\dagger H_u\right|^2 \, .
\ee
Here $g_1$ and $g_2$ are the gauge couplings of $\smash{U(1)_Y}$ and $\smash{SU(2)_L}$, respectively. SUSY breaking in addition induces the soft terms
\be
V_\text{Higgs} \supset m_{H_u}^2 |H_u|^2  + m_{H_d}^2  |H_d|^2 +  m_S^2 |S|^2 -  \left( a_\lambda S  H_u  H_d + \frac{1}{3} a_\kappa S^3 + \text{h.c.} \right) \, .
\ee

The Higgs-singlet coupling in the superpotential contributes to the quartic Higgs coupling. This changes the well-known upper bound on the tree-level mass of the lightest $CP$-even Higgs to (see e.g.~\cite{Ellwanger:2009dp})
\be
m_h^2 \, \leq \, m_Z^2 \left(\cos^2 2 \beta  + \frac{2 \lambda^2}{g_1^2 +g_2^2}\sin^2 2 \beta \right)  ,
\ee
where $m_Z$ is the $Z$-mass. For small $\tan \beta$ and large $\lambda$, the (tree-level) Higgs mass can thus be much larger than without the singlet and can be raised above the LEP bound of $114$ GeV. We therefore focus on this regime in the following. We note that small $\tan \beta \lesssim 3$ is also favoured by electroweak precision tests when the Higgs is heavy \cite{Barbieri:2006bg} (see Sect.~\ref{sec:Constraints} for more details).

How large can $\lambda$ be? In the NMSSM, the size of $\lambda$ is limited by the requirement that the coupling stays perturbative up to the GUT scale. As the Higgs is dual to a composite state in our model, we only need to ensure perturbativity of $\lambda$ at energies below the IR cutoff (corresponding to the compositeness scale). The RG equation for $\lambda$ is
\be
\frac{d \lambda^2}{d \ln \mu} = \frac{\lambda^4}{2 \pi^2} \,,
\ee
 where $\mu$ is the RG scale. For example for $\smash{\lambda=1.8}$ at $\smash{\mu =500 \text{ GeV}}$, the coupling becomes nonperturbative ($\smash{\lambda^2 > 4 \pi}$) around $\mu = 45$ TeV. As we discuss in Sect.~\ref{sec:Constraints}, this is above the IR cutoff that we assume. For such large $\lambda$, the (tree-level) Higgs mass can be as heavy as 300 GeV. Since large loop corrections to the quartic coupling are no longer needed to push the Higgs mass above the LEP bound, this ameliorates the little hierarchy problem of the MSSM. 

Fine-tuning is further reduced because naturalness bounds on soft masses are considerably relaxed when the Higgs is heavy. In particular, allowing 20\% tuning, the naturalness bounds on stops and gluinos are \cite{Barbieri:2006bg}
\be
\begin{split}
m_{\tilde{t}}  & \, \lesssim \, 600 \text{ GeV} \\
M_{3}  &\,  \lesssim \, 1.2 \text{ TeV} \, .
\end{split}
\ee
We will assume that the stops (and the left-handed sbottom) are heavy enough not to be the LSP and that the gluino mass is close to this naturalness bound.  We will furthermore assume that also the electroweak gauginos are relatively heavy as allowed by naturalness. With heavy gauginos there is a large region in parameter space for which a heavy Higgs is compatible with electroweak precision tests (see Sect.~\ref{sec:Constraints} for more details). The mixing of Higgsinos with gauginos is suppressed by the small gauge couplings but their mixing with the singlino (from the superfield $S$) is enhanced by the large coupling $\lambda$. When the gauginos are heavy, we can therefore neglect their mixing with the Higgs sector fermions. In that case, the charged Higgsino has the mass $\smash{\mu_\eff\equiv\lambda\langle S\rangle}$ and the masses of the three neutralinos $\smash{\widetilde{N}_1}$ to $\smash{\widetilde{N}_3}$ fulfill \cite{Barbieri:2006bg}
\be
|m_{\widetilde{N}_1}| \, \leq \,  \mu_\eff \,< \, |m_{\widetilde{N}_2}|,|m_{\widetilde{N}_3}| \, .
\ee 
In particular, the lightest neutralino is the LSP. This Higgsino LSP is a viable dark matter candidate as the mixing with the singlino sufficiently reduces the annihilation cross section \cite{Barbieri:2006bg} (compared to a pure Higgsino LSP for which the relic abundance is too low \cite{Mizuta:1992qp}).  Thus, we have a perfectly acceptable dark matter candidate.

It was shown in \cite{Franceschini:2010qz} that, in a stable minimum and for $\kappa <\lambda$, the parameter $\mu_\eff$ lies in the range
\be
\sqrt{2} \, \frac{\lambda \, m_Z \sin 2 \beta }{\sqrt{g_1^2+g_2^2}}\, \lesssim \; \mu_\eff \; \lesssim \, 3\, \frac{ \lambda \, m_Z  \sin 2 \beta}{\sqrt{g_1^2+g_2^2}} \, .
\ee
For small $\tan \beta$ and large $\lambda$, the lightest chargino is thus sufficiently heavy to avoid the LEP bound on its mass. This solves the $\mu$-problem of the MSSM.

Gravity mediation and anomaly mediation induce $A$-terms which are much smaller than the soft scale $\smash{m_\soft^\ir}$ (see Sect.~\ref{SUSYbrUV}).
In the limit $\smash{a_\lambda,a_\kappa\rightarrow 0}$, the Higgs potential is invariant under the $R$-symmetry of Sect.~\ref{sec:R-symmetry}. The Higgs vevs break this symmetry spontaneously and the Higgs spectrum correspondingly contains a very light pseudo-Nambu-Goldstone boson for $a_\lambda,a_\kappa \ll m_\soft^\ir$. The gaugino masses however contribute radiatively to the $A$-terms and thereby raise the mass of this $R$-axion.\footnote{It can happen that the $R$-axion remains so light that it decays primarily to $\tau \bar{\tau}$. If the Higgs decays dominantly to such $R$-axions, the LEP bound on the Higgs mass is lowered to 90 GeV \cite{Dermisek:2005ar}. This offers an alternative solution to the SUSY little hierarchy problem. } The relevant RG equations are (see e.g. \cite{Ellwanger:2009dp})
\begin{align}
\frac{d a_\lambda}{d\ln \mu}
	& \, =\,  	- \frac{3}{8 \pi^2} \lambda \left[ \frac{1}{5} g_1^2 M_1  \, + \, g_2^2  M_2 \right] \, + \, \cdots
	\\
\frac{d a_\kappa}{d \ln \mu}
	& \, = \,   \frac{3}{8 \pi^2} \left[ \lambda^2 a_\lambda \, + \, 3 \kappa^2 a_\kappa \, + \, 2 \lambda \kappa a_\lambda \right] \, .
\end{align}
Here $M_1$ and $M_2$ are the masses of $\smash{U(1)_Y}$- and $\smash{SU(2)_L}$-gauginos, respectively. Since the log-factor from the RG running, $\smash{\ln(m_\ir/m_\ew)}$, only partially cancels the loop-factor, we expect $\smash{a_\lambda \sim M_2/10 \sim 100 \text{ GeV}}$. This already raises the mass of the $R$-axion sufficiently to avoid collider constraints. The gauginos only contribute at two-loop order to $a_\kappa$ which therefore is much smaller.

\section{The little hierarchy}
\label{sec:LittleHierarchy}
Let us take a closer look at the little hierarchy in our model. To determine the maximal little hierarchy that stops, gauginos and Higgsinos can naturally stabilize, we will discuss various loop corrections to the Higgs sector.

We begin with one-loop corrections due to  standard model fermions. The largest contributions come from the top and the bottom. Top loops are rendered safe by corresponding stop loops. The bottom contribution, on the other hand, is only partially cancelled because the right-handed sbottom is very heavy. But it can still be sufficiently small due to the small Yukawa coupling and since loop corrections to the Higgs are cut off at the warped-down scale $\Lambda_\ir$. Let us assume that $\smash{\tan \beta = \mathcal{O}(1)}$ and that the mass of the (lightest $CP$-even) Higgs is $\smash{250 \text{ GeV}}$. Using Eq.~\eqref{one-loop}, we find that for
\be
\label{cutofflimit}
\Lambda_\ir \, \lesssim \, 170 \text{ TeV} \, ,
\ee 
the bottom-sbottom contribution is less than five times the Higgs mass-squared. This indicates that even with this relatively high cutoff, the tuning required as a consequence of these one-loop corrections can be less than $20\%$.\footnote{See however the caveat in footnote \ref{HiggsCaveat}. We expect a comparable naturalness constraint on $\Lambda_\ir$ if we directly consider the tuning to obtain the correct $Z$-mass. This is in particular the case if the other scalars from the Higgs sector are not too close in mass to the lightest $CP$-even Higgs (see \cite{Barbieri:2010pd2}).} Due to other corrections discussed below, we will however take the much lower cutoff $\smash{\Lambda_\ir \approx 40 \text{ TeV}}$ in the following. We will furthermore choose the KK mass scale $\smash{m_\kk \approx 30 \text{ TeV}}$, so that there is at least one KK mode in each tower before the theory becomes strongly coupled. This corresponds to an IR scale $\smash{m_\ir \approx 10 \text{ TeV}}$.

Next we consider one-loop corrections due to higher KK modes of standard model fermions. Since KK modes are localized in the IR and we assume $5D$ Yukawa couplings of order one (in units of $k$), these fermions couple unsuppressed to the Higgs. The resulting large contributions to the Higgs mass are cancelled by their scalar superpartners, up to an amount that is determined by the mass difference between the superpartners. Scalar KK modes obtain SUSY-breaking masses from gravity-mediation in the bulk and the messengers on the UV brane. The former contribution is of order the soft scale $\smash{m_\soft^\ir}$ but the latter contribution can potentially be larger. To see this, let us determine the wavefunction overlap with the UV brane of a scalar KK mode with mass of order $m_\kk$. Ignoring the gravity-mediated soft mass in the bulk and assuming that the wavefunction is not significantly affected by the soft mass on the UV brane, we find  (see \cite{Gherghetta:2000qt})
\be
\label{wfo}
f(y=0) \, \sim \, \sqrt{k} \, e^{- k \ell \, |c-\frac{1}{2}|} \, ,
\ee
where $c$ is the mass parameter of the corresponding hypermultiplet. For $c$ close to $\smash{\frac{1}{2}}$, the wavefunction overlap is large and we expect that the scalar KK mode picks up a large soft mass from the UV brane. Correspondingly, we expect that the correction to the Higgs mass from such a scalar KK mode and its fermionic superpartner is relatively large.

Let us therefore determine the correction to the Higgs mass due to KK modes and zero-modes from a given hypermultiplet. More precisely, Yukawa couplings to the Higgs involve two hypermultiplets. For simplicity,  we will assume that both hypermultiplets have the same mass parameter $c$, so that their KK decompositions agree. Similarly, we take only the soft mass $\smash{m_\soft^\uv \gtrsim 1000 \text{ TeV}}$ on the UV brane into account but ignore the much smaller gravity-mediated soft mass in the bulk. Before performing a careful calculation using $5D$ propagators, let us estimate the correction taking only modes with masses below the cutoff $\Lambda_\ir$ into account. We then have to distinguish two cases: For
small $c$, the lowest scalar mode (which is massless in the SUSY limit) is highly localized in the IR and correspondingly picks up only a small soft mass from the UV brane. There are then four KK modes\footnote{Here we only count modes with even boundary conditions as odd modes do not couple to the Higgs.} from a given hypermultiplet with masses below  $\smash{\Lambda_\ir=40 \text{ TeV}}$: A massless fermion (the standard model fermion), a light scalar (its superpartner) and a scalar and a fermion with mass of order $m_\kk$. Ignoring log-factors from the loop integrals, the combined correction to the Higgs mass is roughly 
\be
\label{estimate1}
\Delta m_H^2 \, \sim \, \frac{1}{16 \pi^2} \, \left( m_{0,b}^2 \, + \, m_{1,b}^2 \,  - \, m_{1,f}^2  \right)\, ,
\ee
where $\smash{m_{i,b}}$ ($\smash{m_{i,f}}$) denote the mass of the $i$-th scalar (fermionic) KK mode. For larger $c$, on the other hand, the lowest scalar mode becomes less and less localized in the IR and correspondingly picks up a larger mass from the UV brane. Finally, it becomes heavier than $\smash{\Lambda_\ir=40 \text{ TeV}}$ (at $c\simeq0.44$ for $\smash{m_\ir=10 \text{ TeV}}$ and $\smash{m_\soft^\uv=1000 \text{ TeV}}$) and only three KK modes remain in the low-energy spectrum. This means that the quadratic divergence in the Higgs mass (which is cut off at $\Lambda_\ir$) due to the fermionic zero-mode is no longer cancelled for larger $c$. As discussed before, this contribution can nevertheless be small because the fermionic zero-mode becomes localized in the UV for larger $c$ and its wavefunction overlap with the IR brane (and thus its coupling to the Higgs) becomes exponentially suppressed. 
In this regime, the correction to the Higgs mass is roughly
\be
\label{estimate2}
\Delta m_H^2 \, \sim \, \frac{1}{16 \pi^2} \, \left(  m_{1,b}^2 \,  - \, m_{1,f}^2 \, + \, y_c^2 \, \Lambda_\ir^2 \right)\, ,\qquad \text {where} \quad y_c  =   \frac{c- \frac{1}{2}}{e^{(2c-1)k \ell} -1}
\ee
is the $4D$ Yukawa coupling of the fermionic zero-modes to the Higgs. 
The mass quantization condition for the scalar KK modes reads (see \cite{Gherghetta:2000qt,Gherghetta:2003wm})
\be
\frac{ \frac{m_n}{k} J_{c - 1/2}\left( \frac{m_n}{k} \right) \, -\, \frac{(m_\soft^\uv)^2}{2k^2} 
     J_{c + 1/2}\left(\frac{m_n}{k}\right)}{ \frac{m_n}{k} Y_{c - 1/2} \left( \frac{m_n}{k}\right) \, - \, \frac{(m_\soft^\uv)^2}{2k^2} 
     Y_{c + 1/2}\left(\frac{m_n}{k}\right)} \, = \, \frac{  J_{c - 1/2}\bigr( \frac{m_n}{m_\ir} \bigr) }{ Y_{c - 1/2} \bigl( \frac{m_n}{m_\ir}\bigr)} \, ,
     \ee
where $J$ and $Y$ are Bessel functions and we restrict ourselves to $\smash{c>-\frac{1}{2}}$. Via SUSY, the masses of the fermionic KK modes are determined by the same relation with $\smash{m_\soft^\uv}$ set to zero. 
Using these results, we have plotted $\Delta m_H \equiv \sqrt{|\Delta m_H^2|}$ for $\smash{m_\ir=10 \text{ TeV}}$ and $\smash{m_\soft^\uv=1000 \text{ TeV}}$ in Fig.~2.  We see that the correction becomes largest in the transition 
region between the validity of Eqs.~\eqref{estimate1} and \eqref{estimate2}. This is not surprising as the scalar zero-modes have a mass of order $\Lambda_\ir$ in that region but the Yukawa coupling $y_c$ of the fermionic zero-modes to the Higgs is still of order one. This also shows that the contribution from higher KK modes in that region (which could have been large) is at most comparable to the contribution from zero-modes. Finally, note that the dip in Fig.~2 is due to an accidental cancellation in Eq.~\eqref{estimate2}. This feature disappears when we determine the correction more carefully (see below).

\addtocounter{figure}{-1}
\begin{figure}[t]
\centering
\subfigure[Estimate of the one-loop correction $\smash{\Delta m_H}$ to the Higgs mass due to KK modes with masses below the cutoff $\Lambda_\ir$.\label{Fig:MassSplit}\addtocounter{figure}{1}]{
\includegraphics[width=6.9cm]{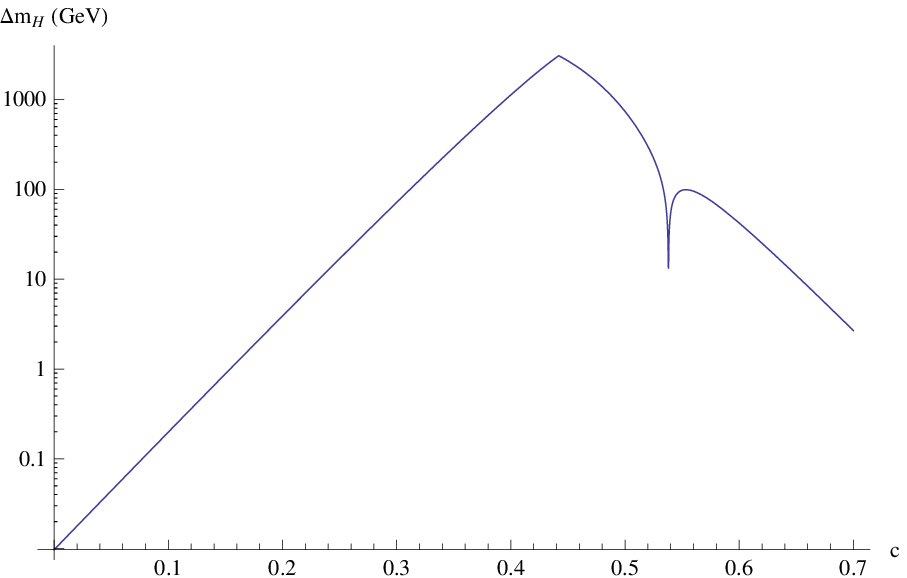} }
\hspace{0.5cm}
\subfigure[One-loop correction $\smash{\Delta m_H}$ to the Higgs mass due to a hypermultiplet using the $5D$ propagator.\label{Fig:LoopCorrection}]{
\includegraphics[width=6.9cm]{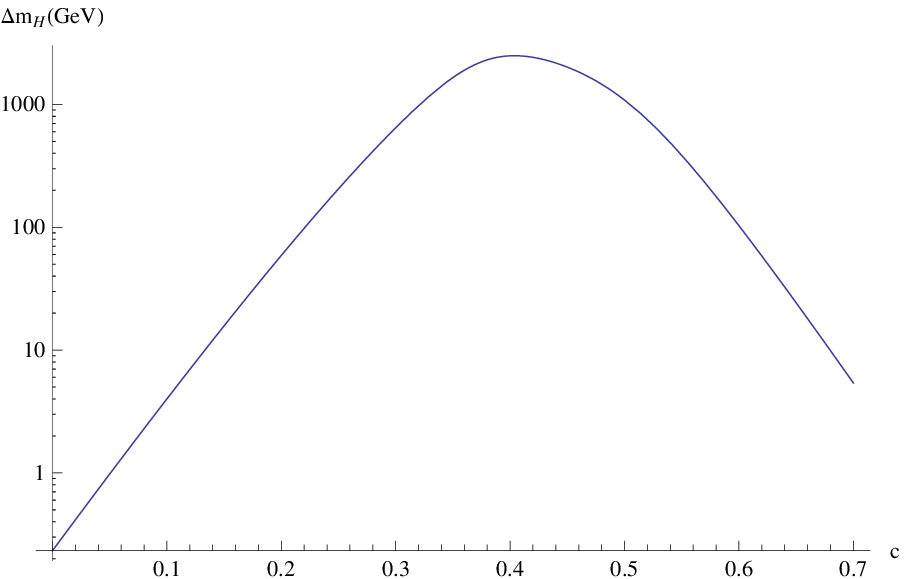} }
\end{figure}

Let us now redetermine the correction to the Higgs mass due to hypermultiplets using $5D$ propagators. This takes the contributions from all KK modes (and the zero-modes) into account. As before, we will assume that both hypermultiplets in the Yukawa coupling to the Higgs have the same $c$-value and we will only take the SUSY-breaking mass on the UV brane into account.\footnote{In absence of SUSY breaking in the bulk, the KK modes effectively lower the cutoff from the new-physics scale $\Lambda_\ir$ to the IR scale $m_\ir$ \cite{Gherghetta:2003wm}. This follows simply from the fact that, with SUSY being only broken on the UV brane, loop corrections on the IR brane are UV-finite by locality and therefore cannot depend on the cutoff $\Lambda_\ir$. It is an interesting question whether the cutoff is still lowered if SUSY is broken in the bulk by the spurion $\Phi$.}
In mixed position-momentum representation and evaluated on the IR brane, the propagators of the fermion and the scalar are given by \cite{Gherghetta:2003wm} 
\be
G_{F,B}(p) \, = \, -  \frac{e^{2 k \ell}}{k} \frac{\widetilde{I}^\uv_{c+1/2}\hspace{-.1cm}\left(\frac{p}{k}\right) \, K_{c+1/2}\bigl(\frac{p}{m_\ir} \bigr) \, - \,
    \widetilde{K}^\uv_{c+1/2}\hspace{-.1cm}\left(\frac{p}{k}\right) \, I_{c+1/2}\bigl(\frac{p}{m_\ir}\bigr) }{\widetilde{I}^\ir_{c+1/2}\hspace{-.02cm}\bigl(\frac{p}{m_\ir}\bigr) \, \widetilde{K}^\uv_{c+1/2}\hspace{-.1cm}\left(\frac{p}{k}\right)  \, - \, \widetilde{I}^\uv_{c+1/2}\hspace{-.1cm}\left(\frac{p}{k}\right) \, \widetilde{K}^\ir_{c+1/2}\hspace{-.02cm}\bigl(\frac{p}{m_\ir}\bigr)  }  \, , 
\ee
where $p$ is the $4D$ momentum and we restrict ourselves to $\smash{c>-\frac{1}{2}}$.  Here $I_\alpha$ and $K_\alpha$ are the modified Bessel functions and $\smash{\widetilde{I}^i_\alpha}$ for $\smash{i \in \{\uv,\ir\}}$  is defined as
\be
\widetilde{I}^i_\alpha(x) \, \equiv  \, 
 x \, I_{\alpha-1}(x) \, - \,  \delta^i \,  I_\alpha(x) \, ,
 \ee
and similarly for $\smash{\widetilde{K}^i_\alpha}$. For the scalar, $\smash{\delta^\uv=(m_\soft^\uv)^2}/2 k^2$ and $\smash{\delta^\ir=0}$ whereas for the fermion $\smash{\delta^\uv=\delta^\ir=0}$. The one-loop correction to the Higgs mass is then given by \cite{Gherghetta:2003wm} 
\be
\Delta m_H^2 \, = \, \frac{3 y_{5D}^2}{4 \pi^2} \int dp \,p^5  \left[G_F^2(p) - G_B^2(p) \right] \, , 
\ee
where $\smash{y_{5D}}$ is the $5D$ Yukawa coupling and we have assumed that the fermionic zero-mode is a quark. In Fig.~3, we have plotted the correction $\Delta m_H = \sqrt{|\Delta m_H^2|}$ for the case $\smash{y_{5D}=k^{-1}}$, $\smash{m_\ir=10 \text{ TeV}}$ and $\smash{m_\soft^\uv=1000 \text{ TeV}}$. Note in particular that our estimate in Fig.~2 reproduces the result in Fig.~3 reasonably well. We see that the correction becomes relatively large for $c$-values close to $0.4$. More precisely, if we want at most $20\%$ tuning for a Higgs mass of $\smash{250 \text{ GeV}}$, we must exclude $c$-values between $0.3$ and $0.53$. We do not expect that this is a problem for successful phenomenology: The statistical analysis performed in \cite{Huber:2003tu} to determine the $c$-values that `most naturally' reproduce the quark masses and the CKM matrix finds two $c$-values which are marginally in that range, $\smash{c=0.317}$ for the top-bottom doublet and $\smash{c=0.528}$ for the right-handed charm.  But we have already discussed in Sect.~\ref{UVcontribution} that there is some freedom in choosing these $c$-values. Note furthermore that in our calculation we have assumed that the left- and right-handed hypermultiplet have the same $c$. We expect that, if only one of both hypermultiplets lies in the region $\smash{0.3<c<0.53}$, the correction will be smaller.

Next, we will discuss loop corrections to the Higgs mass from the gauge sector. Since the SUSY-breaking gaugino masses in the bulk and on the branes are at most of order $\smash{m_\susyb^\ir}$, we can restrict our discussion to the zero-modes from the vector multiplets. 
Light gauginos, however, are not enough to guarantee that these corrections are sufficiently small:
The superpartners of the first two generations obtain high masses $\smash{m_{\soft}^\uv\gtrsim 1000 \text{ TeV}}$ and can be integrated out below that scale. Without their superpartners, the first-two-generation fermions only renormalize the gauge couplings and not the couplings involving gauginos or $D$-terms (at one-loop). This induces a (hard SUSY-breaking) mismatch between these couplings and thereby gives a contribution to scalar masses in the IR \cite{Sundrum:2009gv},
\be
\label{two-loop}
\Delta m_\scalar^2 \, \approx \, \frac{n^2-1}{6 \pi^2 n \gamma_n} \frac{g_n^4}{16 \pi^2} \left[\left(\frac{m_{\soft}^\uv}{\Lambda_\ir}\right)^{\gamma_n} - \, 1\right] \, \Lambda_\ir^2\, ,
\ee
where respectively $\smash{n=2,3}$ for $\smash{SU(2)_L}$ and $\smash{SU(3)_C}$. The parameter $\gamma_n$ is the anomalous dimension of the $D$-term-squared in the dual picture (see \cite{Sundrum:2009gv} for more details).  As in \cite{Sundrum:2009gv}, we will take $\smash{\gamma_2 =1/12}$ and $\smash{\gamma_3 =1/4}$ as well as $\smash{g_2=0.6}$ and $\smash{g_3 =1}$. Choosing $\smash{\Lambda_\ir = 40 \text{ TeV}}$ and $\smash{m_{\soft}^\uv=1000 \text{ TeV}}$, we find the following contributions to stop and Higgs masses:\footnote{In contrast to the findings of \cite{NaturalnessConstraints,ArkaniHamed:1997ab}, these contributions are relatively small even when the first-two-generation superpartners have masses around $1000 \text{ TeV}$. The reason is that our cutoff $\Lambda_\ir$ is low and that the CFT (in the dual picture) provides a focussing effect for the vector multiplet couplings \cite{Sundrum:2009gv}.}
\be
\label{two-loop2}
\Delta m_{\tilde{t}}^2 \, \approx \, (1.5 \text{ TeV})^2 \quad \quad \quad \Delta m_{\text{H}}^2 \, \approx \, (350 \text{ GeV})^2 \, .
\ee
A stop mass $\smash{m_{\tilde{t}}\approx 600 \text{ GeV}}$ (as motivated in Sect.~\ref{secHiggs}) thus requires a tuning in the $20\%$-range of this contribution against the gravity-mediated soft mass. The contribution to the Higgs mass in Eq.~\eqref{two-loop2} in turn is negligible compared to that from stop loops (which again introduce a tuning of order $20\%$ for $\smash{m_{\tilde{t}}\approx 600 \text{ GeV}}$, see Sect.~\ref{secHiggs}). Both contributions in Eq.~\eqref{two-loop2} can be reduced with a lower cutoff or with smaller $\gamma_n$. 

There is a potentially large correction to the Higgs mass from a $D$-term which forces us to extend the gauge group of the standard model as we will now discuss \cite{Strassler:2003ht,Sundrum:2009gv}: Recall first that a bulk vector multiplet consists of a vector multiplet $V$ and a chiral multiplet $\chi$ of $\smash{\mathcal{N}=1}$ SUSY. The $D$-term in $V$ is related to the scalar in $\chi$ according to \cite{Marti:2001iw}
\be
D \, = \, e^{-2 k |y|} \left[\partial_y - 2 k \epsilon(y) \right] \chi \, + \, \cdots \, ,
\ee
where the ellipsis denote contributions from bulk hypermultiplets and brane-localized chiral multiplets. A term linear in $D$ on the UV brane, if induced by broken SUSY, causes the profile of the scalar in $\chi$ to grow rapidly towards the IR. This leads to excessive SUSY breaking in the IR. To ensure the absence of terms linear in $D$, the $D$-terms have to carry a charge. In the standard model, however, the hypercharge $D$-term is not charged under any symmetry. We therefore have to extend the standard model group \cite{MoreMinimal,Sundrum:2009gv}. One possibility is a semi-simple GUT since $D$-terms for nonabelian gauge groups are in the adjoint representation. Another possibility is the left-right model of \cite{Mohapatra:1974gc}\footnote{This model has the gauge group $\smash{SU(3)_C \times SU(2)_L \times SU(2)_R \times U(1)_{B-L}}$ and a discrete parity symmetry. In the simplest realization, one would assign right-handed standard model fields to doublets of $\smash{SU(2)_R}$. Since bulk fields in the same doublet have the same mass parameter $c$, it may however be difficult to obtain the right $4D$ Yukawa couplings without some hierarchy in the $5D$ Yukawa couplings. Alternatively, one may introduce additional particles to fill up the doublets.} in which the abelian $D$-term obtains a discrete charge.\footnote{Similarly, we have to ensure that the $D$-term of the $\smash{U(1)'}$ (see Sect.~\ref{UVcontribution}) has a discrete charge. Alternatively, we can replace the $\smash{U(1)'}$ by a nonabelian gauge symmetry.} This extended symmetry moreover ensures that
\be
\label{D-term}
\Tr \left[Y_i \mathbf{m}^2_{\scalar,i} \right] \, = \, m_{H_u}^2  - m_{H_d}^2 + \Tr \left[ \mathbf{m}^2_Q - \mathbf{m}^2_L  -2 \, \mathbf{m}^2_u + \mathbf{m}^2_d + \mathbf{m}^2_e \right]\, \lesssim \, m_\soft^\ir \, ,
\ee
where $Y_i$ denotes the hypercharge of the $i$-th scalar and the mass matrices of scalar superpartners $\mathbf{m}$ on the right-hand side are $\smash{3\times 3}$-matrices in family space, 
and thereby that $D$-term interactions give sufficiently small contributions to the Higgs mass (see e.g.~\cite{Martin:1997ns}).
In the following, we will assume such an extension of the standard model. We will furthermore assume that the extended group is broken down to the standard model at the IR scale, leading to additional gauge bosons with masses around that scale.

Let us now see how the little hierarchy between $10$ TeV and the electroweak scale, 
\be
\label{RequiredHierarchy}
\frac{m_\ir}{m_\ew} \, \approx \, 40 \, ,
\ee
can be generated in our model. Recall from Eq.~\eqref{msoftIR} that the hierarchy between the IR scale and the soft scale is determined by the quantity $\smash{(C_\ir/M_5)^3}$. The soft scale in turn sets the scale of dimensionful parameters in the Higgs potential and thereby controls the scale of electroweak symmetry breaking. For definiteness, we will set $\smash{m_\soft^\ir = m_{\tilde{t}} \approx 2 m_\ew}$. The little hierarchy in Eq.~\eqref{RequiredHierarchy} then follows from a very modest hierarchy between $M_5$ and $C_\ir$:
\be
\frac{M_5}{C_\ir} \, \approx \, 3.5\, .
\ee

To summarize, stops, gauginos and Higgsinos protect the Higgs from the cutoff down to the electroweak scale. We estimate that the residual tuning required to obtain the correct $Z$-mass for $\smash{\Lambda_\ir \approx 40 \text{ TeV}}$ is in the $20 \%$-range. The little hierarchy between the cutoff and the electroweak scale is controlled by the quantity $\smash{(M_5/C_\ir)^3}$ which can be large already for very modest hierarchies between $M_5$ and $C_\ir$ due to the appearance of the third power.

\section{Electroweak and Flavour Constraints}
\label{sec:Constraints}
Let us briefly discuss various constraints on our model, beginning with electroweak precision tests. Important constraints arise from the Peskin-Takeuchi parameters $S$ and $T$ and the $Zb\bar{b}$-coupling. In warped models with low KK mass scale, a custodial symmetry is required to satisfy these constraints. But for our $\smash{m_\ir \approx 10}$ TeV (see Sect.~\ref{sec:LittleHierarchy}), the KK modes have masses of at least $\smash{m_\kk \approx 30 \text{ TeV}}$. This high scale ensures that corrections to $S,T$ and the $Zb\bar{b}$-coupling from KK modes are sufficiently small even without a custodial symmetry \cite{Davoudiasl:2009cd}. But we similarly have to check whether no excessive corrections arise from the light modes. In particular, we want a heavy Higgs to minimize fine-tuning. For large Higgs-singlet coupling $\lambda$, the Higgs mass can be as large as 300 GeV (cf.~Sect.~\ref{secHiggs}). By itself, such a heavy Higgs would be incompatible with electroweak precision tests. The contributions to $S$ and $T$ from the scalars and fermions in the Higgs sector, stops and sbottoms were determined in \cite{Barbieri:2006bg}, assuming that the gauginos are sufficiently heavy to be neglected (cf.~Sect.~\ref{secHiggs}). It turns out that the contribution  from a heavy Higgs is cancelled by the other Higgs sector particles (whose contributions are enhanced for large $\lambda$) for $\smash{\tan \beta \lesssim 3}$ and that most of the parameter space is brought into the region of the $S-T$ ellipse preferred by experiment. This is another reason why we focus on small $\tan \beta$ in this work.

In addition, important constraints arise from flavour- and $CP$-violating processes. Extensive studies of such processes in warped models with bulk fermions were undertaken in \cite{Agashe:2004cp,Bona:2007vi} (see  \cite{Isidori:2010kg} for a review and additional references). It was found that the strongest constraints arise from the electric dipole moment of the neutron and $CP$ violation in Kaons. In particular, these constraints require respectively that $\smash{m_\kk \gtrsim 6 y_{5d} \text{ TeV}}$ \cite{Agashe:2004cp} and $\smash{m_\kk \gtrsim 49/y_{5d} \text{ TeV}}$ \cite{Isidori:2010kg}, where $y_{5d}$ sets the overall scale of $5D$ Yukawa couplings in units of $k$. The fact that these constraints cannot be fulfilled for low KK mass scales is known as the RS $CP$ problem \cite{Agashe:2004cp}. But assuming that $\smash{y_{5d}\approx 2}$ (which still allows for one or two KK levels before the $5D$ Yukawa couplings become strongly coupled, see e.g.~\cite{Isidori:2010kg}), our KK mass scale of $\smash{30\text{ TeV}}$ is sufficiently high. 

This ensures that KK modes of standard model fermions and gauge bosons do not lead to excessive flavour and $CP$ violation. But in our supersymmetric model, we also have to consider contributions from their superpartners. As we have discussed in the introduction, we avoid the flavour problem of the MSSM
by raising the masses of the first-two-generation sparticles above 1000 TeV. Compared with fermionic and gauge KK modes, higher KK modes of scalar superpartners have two additional sources of flavour and $CP$ violation: 1) Gravity mediation in the bulk leads to flavour-violating masses of order $\smash{m_\soft^\ir}$. 2) Assuming that the messenger sector on the UV brane is not flavour-blind, there are large flavour-violating masses on the UV brane.\footnote{An alternative possibility is a flavour-blind mechanism such as gauge mediation. An $R$-sym-metric version (so that gauginos remain light even when the SUSY-breaking scale is high) of gauge mediation has been presented in \cite{Amigo:2008rc}.} As we have seen in Eq.~\eqref{wfo}, the wavefunction overlap of scalar KK modes with the UV brane can be large, meaning that they may obtain large flavour-violating masses from the UV brane.

Let us assume that flavour is maximally violated by the masses of scalar KK modes.
This is a very conservative assumption as the wavefunction overlap with the UV brane is large
only for $c$ close to $\smash{\frac{1}{2}}$. We will now show that even then flavour- and $CP$-violating processes mediated by scalar KK modes
are sufficiently small. We first
consider $\smash{\Delta F = 2}$ processes which contribute to $B$-$\smash{\overline{B}}$ and
$K$-$\smash{\overline{K}}$ mixing.  The leading RS diagram is tree-level KK gluon
exchange\cite{Agashe:2004cp}, whereas the SUSY diagrams are at loop level.
The SUSY diagrams have the same flavour structure as the tree-level diagrams
and are therefore loop-suppressed compared to the KK gluon exchange; the
tree-level process itself is sufficiently small for $\smash{m_\kk \approx 30 \text{
TeV}}$.  Next are the $\smash{\Delta F = 1}$ processes, such as $\smash{b \to s s s}$, $\smash{b \to
s \ell^+ \ell^-}$, and $\smash{b \to s \gamma}$.  For $\smash{b \to s s s}$ the RS diagrams
are again tree-level while the SUSY diagrams are loop-suppressed.  The
process $\smash{b \to s \ell^+ \ell^-}$ is suppressed relative to the SM diagrams by a factor
$\smash{(\mu_\eff/m_\kk)^2/\abs{V^{\text{CKM}}_{ts}}}$ (where $\mu_\eff$ is the effective $\mu$-term, see Sect.~\ref{secHiggs}).  The 
process $\smash{b \to s \gamma}$, finally, occurs at one-loop in both RS models and SUSY
models; however, the dimension-$5$ operators require a chirality flip which
for SUSY scalars can only originate from $A$-terms or the SUSY Higgs mass
times a Yukawa coupling.  The $A$-terms can never be large in our model as
the Higgs is confined to the IR brane and they are protected by an
approximate $R$-symmetry.  Consequently, the remaining SUSY scalar diagrams
are identical to those of KK gluon exchange \cite{Agashe:2004cp} except that
the scalar diagrams come with a suppression factor of $\smash{\mu_\eff/m_\kk}$.  The
dimension-$6$ operators for $\smash{b \to s \gamma}$ do not require a chirality
flip, but they are also sufficiently small for $\smash{m_\kk \approx 30 \text{ TeV}}$.
An additional concern is from new $CP$ violation---particularly from
contributions to $\epsilon_K$ and the neutron's electric dipole
moment \cite{Agashe:2004cp,Isidori:2010kg}.  For $\smash{m_\kk \approx 30 \text{ TeV}}$, the RS
processes are sufficiently suppressed \cite{Agashe:2004cp,Isidori:2010kg}; meanwhile, as
these $CP$-violating effects require a chirality flip, the SUSY diagrams are
again suppressed relative to the RS graphs by at least $\smash{\mu_\eff/m_\kk}$.

\section{Conclusions}
A warped extra dimension and supersymmetry are popular solutions to the hierarchy problem. In addition, a warped extra dimension allows the fermion mass hierarchy to be naturally generated. Both however are not without issues: In Randall-Sundrum models, the IR scale has to be chosen much larger than the electroweak scale to satisfy phenomenological constraints, leading to a little hierarchy problem. In the MSSM, on the other hand, the tree-level Higgs mass is below the LEP bound. To raise this mass radiatively, heavy stops are required, again causing a little hierarchy problem. Additional issues in the MSSM are the flavour problem of supersymmetry breaking and the $\mu$-problem.\footnote{Of course, various solutions to all these problems exist. But often these solutions are plagued by other problems (the tachyonic-slepton problem in anomaly mediation or the $\smash{\mu/B\mu}$-problem in gauge mediation) or require complicated additional structure (such as flavour symmetries or custodial symmetries in Randall-Sundrum models).} We have shown that the combination of a warped extra dimension and supersymmetry can lead to a model where these problems are naturally avoided. Such a combination seems natural in light of string realizations of the Randall-Sundrum model \cite{Klebanov:2000hb}. 

More precisely, we have shown how to generate and stabilize a little hierarchy between the IR scale and the electroweak scale in a supersymmetric Randall-Sundrum model. This allows for an IR scale around 10 TeV without causing excessive fine-tuning. The resulting Kaluza-Klein scale is sufficiently heavy to comply with all phenomenological constraints (in particular from $CP$ violation) on Randall-Sundrum models without the need to introduce custodial or flavour symmetries. 

To this end, the warped extra dimension is stabilized by a bulk hypermultiplet and a constant superpotential on the IR brane. 
Supersymmetry is broken on the UV brane \cite{Gherghetta:2003wm}, leading to a superparticle spectrum where the first-two generation squarks and sleptons are above 1000 TeV. This  avoids the supersymmetric 
flavour and $CP$ problems. The gaugino is protected by an accidental $R$-symmetry \cite{Sundrum:2009gv} but eventually obtains a mass close to the electroweak scale from either radion mediation or a suppressed operator on the UV brane. The bulk hypermultiplet moreover obtains an $F$-term from the UV brane and induces soft masses near the IR brane for stops, sbottoms and the Higgs sector.
The resulting soft scale can naturally be a factor $10-100$ below the IR scale, thereby generating a little hierarchy between both scales. This little hierarchy is stabilized against radiative corrections by stops, gauginos and Higgsinos, which are the only superpartners in the low-energy spectrum apart from sbottoms and possibly the gravitino. This is reminiscent of the particle content in the more minimal supersymmetric standard model and related scenarios~\cite{Cohen:1996vb,MoreMinimal}.

To alleviate the supersymmetric little hierarchy problem, the Higgs sector on the IR brane is extended with the inclusion of a singlet, as in the NMSSM. This yields an additional contribution to the Higgs quartic coupling and can therefore raise the Higgs mass already at tree-level. Since we have to require perturbativity of the Higgs-singlet coupling only up to the IR scale, this effect can be much larger than in the NMSSM. Tuning is further reduced because the Higgs can be relatively heavy. In addition, log-factors from loop integrals are relatively small due to the low cutoff. We have given an example where the residual tuning to obtain the correct $Z$-mass is in the $20\%$-range. Examples with even less tuning may also be possible. Similar to the NMSSM, the singlet moreover provides a solution to the $\mu$-problem. To avoid a large $D$-term contribution to soft masses the model must be embedded into a GUT group. This allows for the possibility of gauge coupling unification as considered in Ref.~\cite{gcuref}. Finally, the LSP is dominantly Higgsino in a well-motivated region of parameter space. This Higgsino LSP is a viable dark matter candidate.

In the $4D$ dual interpretation the Higgs sector is composite. This protects the Higgs mass from the Planck scale to the IR scale of order 10 TeV. The remaining little hierarchy between the 
IR scale and the electroweak scale is explained by accidental supersymmetry at low energies. Supersymmetry breaking occurs in the elementary sector and is transmitted to the composite sector
by marginal or marginally irrelevant operators. The stabilization of the warped extra dimension requires
a constant IR superpotential which can be thought of as a  gaugino condensate in a Seiberg dual theory.
It is beyond the scope of this work, but it would be worth exploring the $4D$ dual interpretation further.

Let us finally comment on the relevance of our model for the LHC. The low-energy spectrum accessible to the LHC has similarities with other realizations of $\lambda$SUSY \cite{Barbieri:2006bg,Franceschini:2010qz,Kitano:2004zd,Harnik:2003rs}: We expect a relatively heavy Higgs and likely have a Higgsino LSP. The only other light standard model superpartners (apart from the Higgsinos) are stops, sbottoms and gauginos. But we also have specific predictions: Soft masses for the Higgs sector, stops and sbottoms result from gravity mediation at the IR scale of order $\smash{10 \text{ TeV}}$. The dominant source of $R$-breaking soft terms, on the other hand, are the gaugino masses. In particular, $A$-terms are generated mainly radiatively. It would be interesting to study electroweak symmetry breaking under these conditions and to see whether further predictions on the spectrum can be obtained. In addition, a radion (and possibly its fermionic superpartner) around the soft scale is a unique prediction of our model compared to general $\lambda$SUSY scenarios and could be discovered at the LHC.
Higher KK modes and the remaining superpartners, on the other hand, may only be accessible to next-generation colliders.

\section*{Acknowledgments}
We thank Arthur Hebecker, Shrihari Gopalakrishna and Michael G.~Schmidt for helpful discussions. 
This work is supported by the Australian Research Council. TG thanks the SITP at Stanford for 
support and hospitality during the completion of this work. BvH and NS thank SLAC for hospitality
during the final stages of this project.

\appendix

\section{Radion stabilization for generic SUSY-breaking potentials}
\label{App:Pot}

Here we shall discuss the minimization of the radion potential for general SUSY-breaking potentials $U(\Phi,F)$.  We will see that we obtain essentially the same minimum as given in Sect.~\ref{secRadionStabilization}.  

Using the general wavefunctions Eqs.~\eqref{bulksolutions1}--\eqref{bulksolutions3} and the boundary conditions Eq.~\eqref{bc1}, we first find that the wavefunctions take the simpler form
\be
\begin{split}
F \,  = \, 	F_\uv \, e^{(4-\Delta) k|y|} \qquad \quad \widetilde{F} \, \equiv \, 0 \qquad \quad \Phi \, = \, \Phi_\uv  \,  e^{(4-\Delta) k|y|}  
	\\
\widetilde{\Phi} \,  = \, \epsilon(y) \, \frac{ F_\uv^\dagger}{(2 \Delta-6)k} \left(	  \rho^{2\Delta-6} \,   e^{(\Delta-1) k|y|}   \, - \,  e^{(5-\Delta) k |y|} \right) \, .
\label{Eq:eom.phitilde.vev}
\end{split}
\ee
The remaining integration constants $F_\uv$ and $\Phi_\uv$ are determined by the boundary conditions
\begin{equation}
\frac{\partial U}{\partial \Phi_\uv} \, = \, 0  \qquad \quad  \frac{\partial U}{\partial F_\uv} \, = \, F_\uv^\dagger \, \frac{1 \, - \, \rho^{2\Delta-6} }{(\Delta-3) k}  \, .
\label{Eq:boundary.conditions}
\end{equation}

According to the assumption discussed in Sect.~\ref{secRadionStabilization}, each $\Phi$ and $F$ comes with a factor $\smash{k^{-1/2}}$ in the SUSY-breaking potential $U(\Phi,F)$ and the only other scale involved is the SUSY-breaking scale $M_\susyb$. It is convenient to write the SUSY-breaking potential as a dimensionless function $\smash{\widehat{U}\equiv U/M_\susyb^4}$ of the rescaled fields $\smash{\widehat{\Phi}\equiv \Phi/(\sqrt{k} M_\susyb)}$ and $\smash{\widehat{F}\equiv F/(\sqrt{k} M_\susyb^2)}$.  The boundary conditions then take the form
\begin{equation}
\frac{\partial \widehat{U}}{\partial \widehat{\Phi}_\uv} \, = \, 0  \qquad \quad  \frac{\partial \widehat{U}}{\partial \widehat{F}_\uv} \, = \, \widehat{F}_\uv^\dagger \, \frac{1 \, - \, \rho^{2\Delta-6} }{(\Delta-3)}  \, .
\label{Eq:boundary.conditions2}
\end{equation}

We are interested in the potential for $\rho \ll 1$. It is therefore sufficient to solve for $\smash{\widehat{\Phi}_\uv}$ and $\smash{\widehat{F}_\uv}$ as power series in $\rho$. From Eq.~\eqref{Eq:boundary.conditions2}, we conclude that these power series take the form (assuming $\Delta>3$): 
\begin{equation}
\label{expansions}
\begin{split}
\widehat{\Phi}_\uv
	& \, = \,	\widehat{\Phi}_\uv^{(0)} \, + \, \widehat{\Phi}_\uv^{(1)} \, \rho^{2 \Delta - 6}  \, +\,  \mathcal{O}(\rho^{4\Delta-12})
	\\
\widehat{F}_\uv
	& \, =	\, \widehat{F}_\uv^{(0)} \, +\,  \widehat{F}_\uv^{(1)} \, \rho^{2 \Delta - 6} \,+ \,  \mathcal{O}(\rho^{4\Delta-12}) \, .
\end{split}
\end{equation}
For a generic function $\smash{\widehat{U}}$, the dimensionless coefficients $\smash{\widehat{\Phi}_\uv^{(0)}}$ etc. will generically be of order one.
Returning to the unrescaled variables, we see that then the coefficients in the expansion of $\Phi_\uv$ are of order $\smash{\sqrt{k} M_\susyb}$ and those for $F_\uv$ are of order $\smash{\sqrt{k} M_\susyb^2}$. 

To determine the radion potential, we also need to know $\smash{\widetilde{\Phi}_\uv}$. Using Eq.~\eqref{Eq:eom.phitilde.vev} and the power series, we find
\be
\begin{split}
\widetilde{\Phi}_\uv
	 \, & = \,	 \frac{ F_\uv^\dagger }{(6-2 \Delta)k} \left(1 \, -\,  \rho^{2 \Delta-6} \right)
	\\
	 \, & =	 \,  \frac{1}{ (6-2\Delta ) k} \left[ F_\uv^{(0) \, \dagger} \, - \, \left( F_\uv^{(0) \, \dagger}  \,-\,  F_\uv^{(1) \, \dagger} \right) \rho^{2 \Delta - 6} \right] \,+\, \mathcal{O}(\rho^{4\Delta -12}) \, .
\end{split}
\ee
Expanding the SUSY-breaking potential in $\smash{\rho}$ and using Eqs.~\eqref{Eq:boundary.conditions} and \eqref{expansions}, we similarly find
\be
U(  \Phi_\uv ,  F_\uv) \,  =\, U\bigl(\Phi_\uv^{(0)}, F_\uv^{(0)}\bigr)\, +\, \left( \frac{F_\uv^{(0) \, \dagger} F_\uv^{(1)}}{ (\Delta - 3) k} \, \rho^{2\Delta - 6} \, +\,  \text{h.c.} \right) \, + \, \mathcal{O}\left(\rho^{4\Delta -12}\right) \, .
\ee

We now have all the necessary results to evaluate the $4D$ effective potential given in Eq.~\eqref{superfieldform}. To determine the dependence on the scalar superpartner of the radion, we promote the radion $\rho$ to a chiral superfield $\omega$ via analytic continuation into superspace \cite{Luty:2000ec,Goh:2003yr}. Performing the $\smash{d^4 \theta}$-integral, the potential in Eq.~\eqref{superfieldform} depends on the lowest component of this chiral superfield which we again denote by $\omega$. Including the contribution from the constant superpotential on the IR brane in Eq.~\eqref{contribution2}, we find for the $\omega$-dependent part of the potential
\begin{equation}
\label{PotGen}
V_{4}
	\,  \supset \,	  \left( \frac{(\Delta-3)}{2} \, M_U^4
				\,  \omega^{2\Delta - 6}
				\, + \,  \text{h.c.}
		  \right)
		\, + \,  3 \, \frac{ C_\ir^6 }{ M_4^2} \, \abs{\omega}^4
		\, + \,M_\susyb^4 \cdot \mathcal{O}(\omega^{4\Delta -12})  \, , 
\end{equation}
where
\be
M_U^4 \, \equiv  \, \frac{ \abs{F_\uv^{(0)}}^2 + 2 i \, \ima \left(F_\uv^{(0) \, \dagger} F_\uv^{(1)}\right)    }{ (\Delta - 3)^2 k} \, .
\ee
We see that, with the replacements $\smash{M_\susyb^4 \rightarrow |M_U|^4}$ and $\smash{ \gamma\rightarrow \gamma + \arg M_U^4}$, this potential is identical to the potential Eq.~\eqref{PotExp}. Its minimum is accordingly given by Eq.~\eqref{RadionMinimum} with the replacement $\smash{M_\susyb \rightarrow |M_U|}$. Since the coefficients in the expansion of $F_\uv$ are generically of order $\smash{\sqrt{k} M_\susyb^2}$ as argued before, we find that $\smash{|M_U|=\mathcal{O}(M_\susyb)}$. The minimum thus has the same dependence on the SUSY-breaking scale as before. 

Let us now consider the constant part of the $4D$ effective potential. Including the contribution from the constant superpotential on the UV brane in Eq.~\eqref{contribution2}, it reads
\be
\label{ConstantPotential}
V_{4}
	\,  \supset \,	U\bigl(\Phi_\uv^{(0)}, F_\uv^{(0)}\bigr)\, -\, \frac{|F_\uv^{(0)}|^2}{(\Delta-3) k} \, -\,  3 \,  \frac{C_\uv^6}{M_4^2} \, .
\ee
From Eq.~\eqref{Eq:boundary.conditions2}, we expect that $\smash{\widehat{U} \sim 1}$ in the minimum for a generic function $\smash{\widehat{U}}$. The first two terms in Eq.~\eqref{ConstantPotential} are then of order $\smash{M_\susyb^4}$. Assuming that their contribution to the $4D$ effective potential is positive, to tune the cosmological constant to zero we have to choose
\be
C_\uv^3 \, \sim \, M_4 \, M_\susyb^2 \, . 
\ee

The expansion leading to Eq.~\eqref{PotGen} breaks down in the region $\smash{\omega \sim 1}$. We have to check whether the energy density in this region is not smaller than in the minimum Eq.~\eqref{RadionMinimum}. When $\omega \sim 1$, however, the energy density from the constant superpotential on the IR brane is of order $C_\ir^6/M_4^2$. Since we assume that $\smash{M_4 M_\susyb^2 \ll C_\ir^3}$, this is much larger than the energy density in the minimum Eq.~\eqref{RadionMinimum}. We thus find that this minimum is again the global minimum.

\section{A simple messenger sector on the UV brane}
\label{App:Mess}
Here we will present a simple messenger sector on the UV brane. For simplicity, we will only discuss messengers which couple to left-handed squarks (or the corresponding chiral multiplets). There will be corresponding messengers with the right quantum numbers to couple to the remaining squarks and sleptons. More precisely, we consider messengers $X_1$, $\widetilde{X}_1$, 
$X_2$ and $\widetilde{X}_2$ whose charge assignments under the standard model gauge group and the $U(1)'$ are given in Table \ref{tableCA}.
\begin{table}
\begin{center}
 \begin{tabular}{|r|r|r|r|r|}
\hline
 & $SU(3)_C$ & $SU(2)_L$ & $U(1)_Y$ & $U(1)'$\\
\hline
$Q$ & $\mathbf{3}$ & $\mathbf{2}$ & $\smash{\frac{1}{3}}$ & $0$ \\
$\Phi$ & $\mathbf{1}$ &  $\mathbf{1}$ & $0$ & $1$\\
$X_1$ & $\mathbf{1}$ & $\mathbf{1}$ & $0$ & $\smash{-\frac{1}{2}}$ \\
$\widetilde{X}_1$ & $\mathbf{1}$ & $\mathbf{1}$ & $0$ & $\smash{\frac{1}{2}}$ \\
$X_2$ & $\mathbf{\bar{3}}$ & $\mathbf{\bar{2}}$ & $\smash{-\frac{1}{3}}$ & $\smash{\frac{1}{2}}$ \\
$\widetilde{X}_2$ & $\mathbf{3}$ & $\mathbf{2}$ & $\smash{\frac{1}{3}}$ & $\smash{-\frac{1}{2}}$ \\
\hline
\end{tabular}
\caption{Charge assignments for the messengers 
$X_1$, $\widetilde{X}_1$, $X_2$ and $\widetilde{X}_2$, the left-handed 
quark superfields $Q$ and the spurion $\Phi$.\label{tableCA}} 
\end{center}
\end{table}
Furthermore, $\Phi$ is the spurion and we have suppressed the family index for the left-handed quark superfields $Q$. The charge assignments allow for the superpotential
\be
W_X \, = \, Q X_1 X_2 \, + \, \Phi X_1 X_1 \, + \, M_X X_1 \widetilde{X}_1 \, + \, M_X X_2 \widetilde{X}_2 \, ,  
\ee
where $M_X$ is a mass scale. Integrating out the messengers generates the coupling in Eq.~\eqref{UVlc1} (up to a loop factor). The $R$-violating coupling in Eq.~\eqref{UVlc2}, on the other hand, is not generated
as the messenger sector preserves the accidental $R$-symmetry. This follows from the following $R$-charge assignments:
\be
\left[X_1\right] \, = \, \left[X_2\right] \, = \, \left[\Phi\right] \, = \, \frac{2}{3} \quad \quad \quad [\widetilde{X}_1] \, = \, [\widetilde{X}_2 ] \, = \, \frac{4}{3}  \, .
\ee
It is straightforward to include messengers for the other squarks and sleptons in a way that either preserves or breaks the accidental $R$-symmetry.


\begin{thebibliography}{99}

\bibitem{Randall:1999ee}
  L.~Randall and R.~Sundrum,
  Phys.\ Rev.\ Lett.\  {\bf 83} (1999) 3370
  [arXiv:hep-ph/9905221].

\bibitem{Grossman:1999ra}
  Y.~Grossman and M.~Neubert,
  Phys.\ Lett.\  B {\bf 474} (2000) 361
  [arXiv:hep-ph/ 9912408].

\bibitem{Gherghetta:2000qt}
  T.~Gherghetta and A.~Pomarol,
  Nucl.\ Phys.\  B {\bf 586} (2000) 141
  [arXiv:hep-ph/0003129].

\bibitem{Huber:2000ie}
  S.~J.~Huber, Q.~Shafi,
  Phys.\ Lett.\  {\bf B498}, 256-262 (2001).
  [hep-ph/0010195].

\bibitem{Agashe:2004cp}
K.~Agashe, G.~Perez and A.~Soni,
  Phys.\ Rev.\ Lett.\  {\bf 93} (2004) 201804
  [arXiv:hep-ph/0406101];
  K.~Agashe, G.~Perez and A.~Soni,
  Phys.\ Rev.\  D {\bf 71} (2005) 016002
  [arXiv:hep-ph/0408134].


\bibitem{Bona:2007vi}
  M.~Bona {\it et al.}  [UTfit Collaboration],
  JHEP {\bf 0803} (2008) 049
  [arXiv:0707.0636 [hep-ph]];
  K.~Agashe {\it et al.},
  Phys.\ Rev.\  D {\bf 76} (2007) 115015
  [arXiv:0709.0007 [hep-ph]];
C.~Csaki, A.~Falkowski and A.~Weiler,
  JHEP {\bf 0809} (2008) 008
  [arXiv:0804.1954 [hep-ph]].


\bibitem{Isidori:2010kg}
  G.~Isidori, Y.~Nir and G.~Perez,
  arXiv:1002.0900 [hep-ph] and references therein.

\bibitem{flavoursymmetries}
A.~L.~Fitzpatrick, G.~Perez and L.~Randall,
  arXiv:0710.1869 [hep-ph];
A.~L.~Fitzpatrick, L.~Randall and G.~Perez,
  Phys.\ Rev.\ Lett.\  {\bf 100} (2008) 171604;
C.~Csaki, A.~Falkowski and A.~Weiler,
  Phys.\ Rev.\  D {\bf 80} (2009) 016001
  [arXiv:0806.3757 [hep-ph]];
J.~Santiago,
  JHEP {\bf 0812} (2008) 046
  [arXiv:0806.1230 [hep-ph]];
C.~Csaki, G.~Perez, Z.~Surujon and A.~Weiler,
  Phys.\ Rev.\  D {\bf 81} (2010) 075025
  [arXiv:0907.0474 [hep-ph]];
M.~C.~Chen, K.~T.~Mahanthappa and F.~Yu,
  Phys.\ Rev.\  D {\bf 81} (2010) 036004
  [arXiv:0907.3963 [hep-ph]].


\bibitem{Barbieri:2000gf}
  R.~Barbieri and A.~Strumia,
  arXiv:hep-ph/0007265.

\bibitem{ArkaniHamed:2001nc}
  N.~Arkani-Hamed, A.~G.~Cohen and H.~Georgi,
  Phys.\ Lett.\  B {\bf 513} (2001) 232
  [arXiv:hep-ph/0105239].


\bibitem{Cohen:1996vb}
  M.~Dine, A.~Kagan and S.~Samuel,
  Phys.\ Lett.\  B {\bf 243} (1990) 250;
  M.~Dine, R.~G.~Leigh and A.~Kagan,
  Phys.\ Rev.\  D {\bf 48} (1993) 4269
  [arXiv:hep-ph/9304299];
  P.~Pouliot and N.~Seiberg,
  Phys.\ Lett.\  B {\bf 318} (1993) 169
  [arXiv:hep-ph/9308363];
A.~Pomarol and D.~Tommasini,
  Nucl.\ Phys.\  B {\bf 466} (1996) 3
  [arXiv:hep-ph/9507462];
  R.~Barbieri, G.~R.~Dvali and L.~J.~Hall,
  Phys.\ Lett.\  B {\bf 377} (1996) 76
  [arXiv:hep-ph/9512388];
 R.~Barbieri, L.~J.~Hall and A.~Romanino,
  Phys.\ Lett.\  B {\bf 401} (1997) 47
  [arXiv:hep-ph/9702315].

\bibitem{MoreMinimal}
 A.~G.~Cohen, D.~B.~Kaplan and A.~E.~Nelson,
  Phys.\ Lett.\  B {\bf 388} (1996) 588
  [arXiv:hep-ph/9607394];


\bibitem{NaturalnessConstraints}
  R.~Barbieri and G.~F.~Giudice,
  Nucl.\ Phys.\  B {\bf 306} (1988) 63;
  S.~Dimopoulos and G.~F.~Giudice,
  Phys.\ Lett.\  B {\bf 357} (1995) 573
  [arXiv:hep-ph/9507282].

\bibitem{ArkaniHamed:1997ab}
  N.~Arkani-Hamed, H.~Murayama,
  Phys.\ Rev.\  {\bf D56 } (1997)  6733-6737.
  [hep-ph/9703259].
  
  
\bibitem{Barbieri:2010pd1}
G.~F.~Giudice, M.~Nardecchia and A.~Romanino,
  Nucl.\ Phys.\  B {\bf 813} (2009) 156
  [arXiv:0812.3610 [hep-ph]];
  
\bibitem{Barbieri:2010pd2}
  R.~Barbieri, E.~Bertuzzo, M.~Farina, P.~Lodone and D.~Pappadopulo,
  JHEP {\bf 1008} (2010) 024
  [arXiv:1004.2256 [hep-ph]].

\bibitem{Gherghetta:2003wm}
  T.~Gherghetta and A.~Pomarol,
  Phys.\ Rev.\  D {\bf 67} (2003) 085018
  [arXiv:hep-ph/0302001].

  \bibitem{Sundrum:2009gv}
  R.~Sundrum,
  JHEP {\bf 1101} (2011)  062.
  [arXiv:0909.5430 [hep-th]].

\bibitem{gbl}
A.~Kehagias and K.~Tamvakis,
  Phys.\ Lett.\  B {\bf 504} (2001) 38
  [arXiv:hep-th/0010112];
B.~Batell and T.~Gherghetta,
  Phys.\ Rev.\  D {\bf 73} (2006) 045016
  [arXiv:hep-ph/0512356];
  T.~Gherghetta and B.~von Harling,
  JHEP {\bf 1004} (2010) 039
  [arXiv:1002.2967 [hep-ph]].

\bibitem{splitSUSY1}
  N.~Arkani-Hamed and S.~Dimopoulos,
  JHEP {\bf 0506} (2005) 073
  [arXiv:hep-th/0405159].

\bibitem{splitSUSY2}
N.~Arkani-Hamed, S.~Dimopoulos, G.~F.~Giudice and A.~Romanino,
  Nucl.\ Phys.\  B {\bf 709} (2005) 3
  [arXiv:hep-ph/0409232].

\bibitem{Redi:2010yv}
  M.~Redi and B.~Gripaios,
  JHEP {\bf 1008} (2010) 116
  [arXiv:1004.5114 [hep-ph]].

\bibitem{Barbieri:2006bg}
  R.~Barbieri, L.~J.~Hall, Y.~Nomura and V.~S.~Rychkov,
  Phys.\ Rev.\  D {\bf 75} (2007) 035007
  [arXiv:hep-ph/0607332].


\bibitem{Franceschini:2010qz}
  R.~Franceschini and S.~Gori,
  arXiv:1005.1070 [hep-ph].

\bibitem{Kitano:2004zd}
  R.~Kitano, G.~D.~Kribs and H.~Murayama,
  Phys.\ Rev.\  D {\bf 70} (2004) 035001
  [arXiv:hep-ph/0402215];
  L.~Cavicchia, R.~Franceschini and V.~S.~Rychkov,
  Phys.\ Rev.\  D {\bf 77} (2008) 055006
  [arXiv:0710.5750 [hep-ph]];
J.~Cao and J.~M.~Yang,
  Phys.\ Rev.\  D {\bf 78} (2008) 115001
  [arXiv:0810.0989 [hep-ph]];
  P.~Lodone,
  JHEP {\bf 1005} (2010) 068
  [arXiv:1004.1271 [hep-ph]].

  

\bibitem{Harnik:2003rs}
  R.~Harnik, G.~D.~Kribs, D.~T.~Larson and H.~Murayama,
  Phys.\ Rev.\  D {\bf 70} (2004) 015002
  [arXiv:hep-ph/0311349];
  A.~Birkedal, Z.~Chacko and Y.~Nomura,
  Phys.\ Rev.\  D {\bf 71} (2005) 015006
  [arXiv:hep-ph/0408329].

\bibitem{Altendorfer:2000rr}
  R.~Altendorfer, J.~Bagger and D.~Nemeschansky,
  Phys.\ Rev.\  D {\bf 63} (2001) 125025
  [arXiv:hep-th/0003117].

\bibitem{Goh:2003yr}
  H.~S.~Goh, M.~A.~Luty and S.~P.~Ng,
  JHEP {\bf 0501} (2005) 040
  [arXiv:hep-th/ 0309103].

\bibitem{Marti:2001iw}
  D.~Marti and A.~Pomarol,
  Phys.\ Rev.\  D {\bf 64} (2001) 105025 
  [arXiv:hep-th/ 0106256].
  
  \bibitem{Luty:2000ec}
  M.~A.~Luty and R.~Sundrum,
  Phys.\ Rev.\  D {\bf 64} (2001) 065012
  [arXiv:hep-th/ 0012158].

\bibitem{Abel:1995wk}
  S.~A.~Abel, S.~Sarkar and P.~L.~White,
  Nucl.\ Phys.\  B {\bf 454} (1995) 663
  [arXiv:hep-ph/9506359].

\bibitem{Abel:1996cr}
  S.~A.~Abel,
  Nucl.\ Phys.\  {\bf B480 } (1996)  55-72.
  [hep-ph/9609323];
  C.~Panagio-takopoulos and K.~Tamvakis,
  Phys.\ Lett.\  B {\bf 446} (1999) 224
  [arXiv:hep-ph/9809475];

\bibitem{Huber:2003tu}
  S.~J.~Huber,
  Nucl.\ Phys.\  B {\bf 666} (2003) 269
  [arXiv:hep-ph/0303183].

\bibitem{Chacko:2000fn}
  Z.~Chacko and M.~A.~Luty,
  JHEP {\bf 0105} (2001) 067
  [arXiv:hep-ph/0008103].

\bibitem{Luty:2002ff}
  M.~A.~Luty,
  Phys.\ Rev.\ Lett.\  {\bf 89} (2002) 141801
  [arXiv:hep-th/0205077].

\bibitem{Ellwanger:2009dp}
  U.~Ellwanger, C.~Hugonie and A.~M.~Teixeira,
  Phys.\ Rept.\  {\bf 496} (2010) 1
  [arXiv:0910.1785 [hep-ph]].


\bibitem{Mizuta:1992qp}
  S.~Mizuta, M.~Yamaguchi,
  Phys.\ Lett.\  {\bf B298 } (1993)  120-126.
  [hep-ph/9208251].

\bibitem{Dermisek:2005ar}
  R.~Dermisek, J.~F.~Gunion,
  Phys.\ Rev.\ Lett.\  {\bf 95 } (2005)  041801.
  [hep-ph/0502105];
  R.~Dermisek, J.~F.~Gunion,
  Phys.\ Rev.\  {\bf D75 } (2007)  075019.
  [hep-ph/0611142].

\bibitem{Strassler:2003ht}
  M.~J.~Strassler,
  arXiv:hep-th/0309122.

\bibitem{Mohapatra:1974gc}
  R.~N.~Mohapatra and J.~C.~Pati,
  Phys.\ Rev.\  D {\bf 11} (1975) 2558;
  G.~Senjanovic and R.~N.~Mohapatra,
  Phys.\ Rev.\  D {\bf 12} (1975) 1502.

\bibitem{Martin:1997ns}
  S.~P.~Martin,
  In *Kane, G.L. (ed.): Perspectives on supersymmetry* 1-98.
  [hep-ph/9709356].

\bibitem{Davoudiasl:2009cd}
  H.~Davoudiasl, S.~Gopalakrishna, E.~Ponton and J.~Santiago,
  New J.\ Phys.\  {\bf 12} (2010) 075011
  [arXiv:0908.1968 [hep-ph]] and references therein.

\bibitem{Amigo:2008rc}
  S.~D.~L.~Amigo, A.~E.~Blechman, P.~J.~Fox and E.~Poppitz,
  JHEP {\bf 0901} (2009) 018
  [arXiv:0809.1112 [hep-ph]].

\bibitem{Klebanov:2000hb}
  I.~R.~Klebanov, M.~J.~Strassler,
  JHEP {\bf 0008}, 052 (2000).
  [hep-th/0007191];
  S.~B.~Giddings, S.~Kachru, J.~Polchinski,
  Phys.\ Rev.\  {\bf D66}, 106006 (2002).
  [hep-th/0105097].

\bibitem{gcuref}
  T.~Gherghetta,
  Phys.\ Rev.\  {\bf D71}, 065001 (2005).
  [hep-ph/0411090]; K.~Agashe, R.~Contino, R.~Sundrum,
  Phys.\ Rev.\ Lett.\  {\bf 95}, 171804 (2005).
  [hep-ph/0502222].

\end{thebibliography}
\end{document}